\documentclass[]{aa}
\usepackage{graphicx, amsmath, color, mathrsfs,multirow,makecell}
\usepackage[utf8]{inputenc}
\graphicspath{{figures/},{figures/},{./}}
\usepackage{txfonts}
\usepackage{natbib,twoopt}
\usepackage[breaklinks=true]{hyperref} 
\bibpunct{(}{)}{;}{a}{}{,}             
\makeatletter

\newcommandtwoopt{\citeads}[3][][]{\href{http://adsabs.harvard.edu/abs/#3}
        {\def\hyper@linkstart##1##2{}
        \let\hyper@linkend\@empty\citealp[#1][#2]{#3}}}

\newcommandtwoopt{\citepads}[3][][]{\href{http://adsabs.harvard.edu/abs/#3}
        {\def\hyper@linkstart##1##2{}
        \let\hyper@linkend\@empty\citep[#1][#2]{#3}}}

\newcommandtwoopt{\citetads}[3][][]{\href{http://adsabs.harvard.edu/abs/#3}
        {\def\hyper@linkstart##1##2{}
        \let\hyper@linkend\@empty\citet[#1][#2]{#3}}}

\makeatother
\usepackage{hyperref}
\usepackage{hypcap}
\definecolor{cyan}{cmyk}{1.,0.,0.,0.2}
\definecolor{vert}{cmyk}{0.5,0.,0.5,0.5}
\definecolor{magenta}{cmyk}{0.,1.,0.,0.1}
\definecolor{verdatre}{cmyk}{0.5,0.,0.5,0.5}
\definecolor{vert_clair}{cmyk}{0.5,0.,0.5,0.2}
\definecolor{yellow}{cmyk}{0.,0.,1.,0.0}
\definecolor{yellow_1}{cmyk}{0.,0.,0.5,0.0}
\definecolor{rouge}{cmyk}{0.,0.4,0.6,0.0}
\definecolor{orange}{cmyk}{0.,0.5,0.5,0.05}
\definecolor{violet}{rgb}{0.5,0.,0.5}
\definecolor{darwin_box}{rgb}{0.988,0.878,0.77}
\definecolor{darwin_text}{rgb}{0.1,0.07,0.02}
\newcommand{\beq}{\begin{equation}}
\newcommand{\eeq}{\end{equation}}
\newcommand{\bea}{\begin{eqnarray}}
\newcommand{\ena}{\end{eqnarray}}

\begin{document}

\title{Stable laws and cosmic ray physics}
\author{Y.~Genolini\thanks{Contact author: yoann.genolini@lapth.cnrs.fr}  \and P.~Salati \and P.~D.~Serpico \and R.~Taillet
}

\institute{
LAPTh, Universit\'e Savoie Mont Blanc \& CNRS, 9 Chemin de Bellevue, B.P.110 Annecy-le-Vieux, F-74941, France
}
%

\date{Received; accepted\\
Preprint numbers : LAPTH-045/16}
%

%
%
\abstract
%
{In the new ``precision era'' for cosmic ray astrophysics, scientists making theoretical predictions cannot content themselves with average trends, but need to
correctly take into account intrinsic uncertainties. The space-time discreteness of the cosmic ray sources, together with a substantial ignorance
of their precise epochs and locations (with the possible exception of the most recent and close ones) play an important role in this sense.
}
%
%
{We elaborate a statistical theory to deal with this problem, relating the composite probability $P(\Psi)$ to obtain a flux $\Psi$ at the Earth
and the single-source probability $p(\psi)$ to contribute with a flux $\psi$. The main difficulty arises from the fact that $p(\psi)$ is a ``heavy tail'' distribution, characterized by power-law or broken power-law behavior up to very large fluxes, for which the central limit theorem does not hold, and leading to distributions different from Gaussian. The functional form of the distribution for the aggregated flux is nonetheless unchanged by its own convolution, that is, it belongs to the so-called {\it stable laws} class.
}
%
{We analytically discuss the regime of validity of the stable laws associated
with the distributions arising in cosmic ray astrophysics, as well as the limitations to the treatment imposed by causal considerations and
partial source catalog knowledge. We validate our results with extensive Monte Carlo simulations, for different regimes of propagation parameters
and energies.}
%
%
{
We find that relatively simple recipes provide a satisfactory description of the probability $P(\Psi)$. We also find that a naive Gaussian fit to simulation
results would underestimate the probability of very large fluxes, that is, several times above the average, while overestimating the probability of
relatively milder excursions. At large energies, large flux fluctuations are prevented by causal considerations, while at low energies, a partial knowledge
of the recent and nearby population of sources plays an important role.
A few proposals have been recently discussed in the literature  to account for spectral breaks reported in cosmic ray data in terms of local contributions. We apply our newly developed theory to assess their probabilities, finding that they are relatively small, typically at the 0.1\% level or smaller, never exceeding 1\%.
}
%
%
{The use of heavy tail distributions is relevant in assessing how likely a  measured cosmic ray flux is to depart from the average expectation in a given model. The existing mathematical theory leading to stable laws can be adapted to the case of interest via some recipes that closely reproduce numerical simulations and are relatively easy to implement.
}

\maketitle
%
%
\section{Introduction}

A striking property of the (Galactic) cosmic ray spectra is their power-law behavior over many
decades in energy. Power laws are omnipresent in nature (as well
as in many domains of interest for human activities), typically associated with self-similar/scale invariant phenomena.
For example, the photon spectra, thought to be tracers of the parent charged cosmic ray particle spectra, of most sources in high-energy astrophysics (supernova remnants, pulsar wind nebulae, diffuse radiations, etc.) at different wavelengths from radio to gamma-rays are also well described by (broken) power laws, typically with spectral index $\alpha$ in the range $-3<\alpha<-1$.
Even the interstellar magnetic field turbulence is well-described by a power-law power spectrum, which in turn justifies
a power-law behavior for the rigidity dependence of the diffusion coefficient describing cosmic ray propagation in the Galaxy.

Not surprisingly, then, the customary calculation of the cosmic ray flux at the Earth postulates a universal power-law source term,
further assuming the limit of a continuous distribution of sources, both spatially and temporally, in a volume modeling the Galaxy: often, a simple cylinder or an effectively infinite slab. This is an approximation, as the distribution of cosmic
ray sources is actually discrete. This approximation is valid if the distribution of
sources is sufficiently dense in space and in time to be described as a continuum, much like in thermodynamics when
the details of microscopic states corresponding to a given macroscopic state are ignored.
This approximation is well-suited for describing the average expectations for the flux at Earth associated to a given hypothesis on propagation
parameters, source spectra, and energetics. However, the discrete nature of cosmic ray sources should
affect other observables, for instance, the time variation of the flux measured, which is nonetheless less important for all but a few phenomenological
consequences.

Indeed, recently, several experiments have established that the cosmic ray spectrum, even within the three decades of energy
probed by a single experiment such as AMS-02, cannot be satisfactorily described by a single power-law (for a recent review see Sec. 2 in~\cite{Serpico:2015caa}). This is observed for protons, helium
nuclei, and some heavier species. Several articles have attempted to explain this phenomenology as the result of the discrete nature of
cosmic ray sources in space and time (as reviewed in Sec. 3.3 of~\citet{Serpico:2015caa} and also in \citet{2012A&A...544A..92B}). While specific scenarios fitting the data can be found, the likelihood of these solutions, given our
statistical knowledge of the source distribution and rate, is unknown.
This is related to the conceptual difficulty that the probability distribution function $p$ of a single source contributing a flux $\psi$ at Earth is not a Gaussian function,
but rather a heavy-tail/power law distribution (see \citet{2012A&A...544A..92B}, and arguments leading to Eqs.~21 and 24 in Sec. 3.2) which, if extrapolated to very large fluxes, does not even have a finite variance; the central limit theorem for the sum flux
probability distribution $P(\Psi)$ cannot be applied. Ironically, the same mathematical property that makes a phenomenological
description of the cosmic ray observables so simple (``power laws'') raises difficulties in theoretical probabilistic assessments starting from $p(\psi)$.
While this problem has been overtly recognized \citep{lee1979statistical,lagutin1995fluctuations}, the quantitative consequences of this fact for cosmic ray physics are still a matter of debate. Far from
being academic, the problem also has important implications for the theoretical limitations in the extraction of propagation parameters, as well as comparing
theoretical models up to current experimental precision. Ultimately, barring an unrealistic, fully deterministic model of the sources of cosmic rays, theoretical predictions
in this field are intrinsically statistical in nature.
In fact, theoretical calculations concerning ultra-high-energy cosmic ray observables have already routinely used ensemble techniques for estimating the flux and mass
composition uncertainties (see e.g., \citeads{2013PhRvD..87b3004A}). Important differences arise, however, due to the quasi-ballistic propagation and the non-negligible loss-effects ruling the extragalactic propagation regime.

Here, we aim at establishing a systematic theory to evaluate these effects for Galactic cosmic rays. Our main focus is on analytical treatment, but all major conclusions
are validated by extensive numerical simulations. By itself, the numerical part of our work represents a major novel contribution to the understanding
of the consequence of the discrete nature of sources for cosmic ray phenomenology. Also, novel when compared to previous publications in this domain, we analyze the effects of
physical limitations on idealized mathematical extrapolations, such as the eventual failure of diffusion equations to cope with constraints such as causality or a priori knowledge about the discrete source distribution.Whenever a comparison with observations is performed, we refer to the proton flux, which is the best measured one due to high statistics, and for which the effects discussed here should be most prominent. Still, the same formalism can apply to
any nuclear species, and could also be extended to leptons (at least qualitatively; for a pioneering earlier investigation of the lepton channel, see~\citet{Mertsch:2010fn}).

As an application, we then focus on the compatibility of
several experimental results on interstellar protons (such as AMS-02, Pamela,
CREAM) with our current understanding of cosmic ray propagation,
properly taking the theoretical uncertainty into account.

This paper is organized as follows. In Sec.~\ref{problem}, we describe the problem and introduce the main difficulties that make
its solution challenging, as well as our strategy to approach them. Section~\ref{technical} deals with the technical aspects of the analytical treatment proposed,
including a discussion of the limitations of an overly naive approach. The reader uninterested by these subtleties and mostly interested in the validation of our description and the phenomenological implications of our work can quickly gloss over this part and focus on the following sections. In Section~\ref{simulations}, we compare our theory with extensive simulations: this allows us to validate the theory, better defining its regimes of validity,  also clarifying when one should be able to draw conclusions relying solely on analytical arguments without the need for long computing time.  A few applications to cases of phenomenological interest are reported in Section~\ref{applications}.
Finally, in Section~\ref{conclusions}, we conclude.
\section{Description of the problem}\label{problem}
Galactic cosmic rays are accelerated at discrete sources, the position and age of which is not known individually,
only statistically. From the statistical distribution of these sources, we can infer the statistical distribution of the cosmic ray flux in
the solar neighborhood. In this section, we describe the issues raised by this program.
The cosmic ray (CR) flux $\Psi$ obeys a diffusion equation which, in its simplest incarnation boils down to the form (energy dependence implicit):

\begin{equation}
\left[\frac{\partial} {\partial t}-K\,\nabla^2\right]\Psi={\cal Q}\,,\label{DiffEq1}
\end{equation}
where $K$ is the spatial diffusion coefficient supposed to be uniform in the diffusion volume and ${\cal Q}$ is the source term. Additional terms accounting for convection, reacceleration, and energy losses are of little relevance for the species and the energy ranges of interest in the following and can be anyway dealt with at very least with standard techniques like the ones implemented in numerical codes such as GALPROP or DRAGON or semi-numerical ones such as USINE.
The solution to this equation can be formally written as an integral over the Galactic volume and over the past Galactic history as:
\begin{equation}
\Psi({\bf x},t)=\int_{-t_{\rm MW}}^t d t_S\int_{V_{\rm MW}} d{\bf x}_S\; {\cal Q}({\bf x}_S,t_S)\;{\cal G}_{B}({\bf x}_S,t_S\to {\bf x},t)\,,
\end{equation}
where $V_{\rm MW}$ is the volume of the Galaxy, given by $V_{\rm MW}\simeq2\,h\,\pi\, R^2$ if the Galaxy is modeled as a cylinder of radius $R$ and half-thickness $h$, of age $t_{\rm MW}$,  and ${\cal G}_B$ is the Green function obtained by solving the analog of Eq.~(\ref{DiffEq1}) with a temporal and spatial delta at the RHS, and the appropriate boundary conditions.
As a reminder, the Green function of the pure diffusive problem (without boundary) with delta centered at spatial position $\boldsymbol{x_S}$ and time $t_S$ would simply write
\begin{equation}
{\cal G}(d=|\boldsymbol{x}-\boldsymbol{x_S}|,\tau=t-t_S)=\frac{1}{(4\,\pi\,K\,\tau)^{3/2}}\exp{\left(-\frac{d^2}{4\,K\,\tau}\right)}\:.
\label{eq:prop3D}
\end{equation}
There are a few issues related to the computation of the CR flux, which we now briefly describe.

First, the source term $ {\cal Q}$ is expected to be a sum of discrete injection points in space-time (the spatial and time
scale of likely accelerators being assumed much shorter than propagation length and time), whose actual positions and epochs are unknown. This is the so-called myriad model approach \citepads{2003ApJ...582..330H} where cosmic rays are sourced by a constellation of point-like objects and not by a continuous jelly. It would then be formally correct to write
\begin{equation}
{\cal Q}({\bf x}_S,t_S)=\sum_i^N \;q_i \;\delta(\boldsymbol{x_i}-\boldsymbol{x}_S) \;\delta(t_i-t_S)\,,
\end{equation}
leading to \begin{equation}
\Psi({\bf x},t)=\sum_i^N \;\psi_i=\sum_i^N \;q_i\;
{\cal G}_{B}({\bf x}_i,t_i\to {\bf x},t)\,.\label{superpositionsource}
\end{equation}
Commonly  Eq.~(\ref{DiffEq1}), sourced by a discrete sum, is replaced with a continuous proxy corresponding to its ensemble average. Specifically, the source term writes
(for simplicity we assume a unique source term, i.e., $q_i=q$):
\begin{align}
\langle {\cal Q}({\bf x}_S,t_S)\rangle&=\left\langle\sum_i^N \;q_i \;\delta(\boldsymbol{x_i}-\boldsymbol{x}_S) \;\delta(t_i-t_S)\right\rangle \nonumber\\ &\simeq \frac{q\,\nu }{V_{\rm MW}}\;\times\;\begin{cases}\Theta(h-|z|)\,\Theta(R-r) &\text{in 3D, or} \\&\\ 2h\;\delta(z)\,\Theta(R-r)&\text{in 2D,}\end{cases}
\label{eq:source_distribution}
\end{align}
where $q$ is the single source spectrum (particles per unit energy) and $\nu$ the source rate per unit time. For example, assuming supernova remnants to be the sources of CRs, a rate of three explosions per century is reasonable. In the second equality, we assume a homogeneous distribution of sources lying within a cylindrical approximation of the Galaxy with radius $R$. Along the vertical direction, the sources are either uniformly distributed inside a disk with half-thickness $h$ (3D case), or pinched inside an infinitesimally thick disk (2D case). The generalization to a different distribution is straightforward.

This source term leads to the theoretical average flux:
\begin{equation}
\langle \Psi({\bf x},t)\rangle=\int_{-t_{\rm MW}}^t d t_S\int_{V_{\rm MW}} d{\bf x}_S\; \langle{\cal Q}({\bf x}_S,t_S)\rangle\;{\cal G}_{B}({\bf x}_S,t_S\to {\bf x},t)\,.
\label{eq:psi_th}
\end{equation}
This is only true on average. We expect the flux observed at the Earth to be ruled by a probability distribution function (pdf) $P(\Psi)$. This function also depends on the actual value of $N$ of the underlying discrete sources. The average flux becomes
\begin{equation}
\langle \Psi \rangle=\int_0^\infty d\Psi\, \Psi\, P(\Psi)\,.\label{AvFluxProb1}
\end{equation}
Obtaining the probability $P(\Psi),$ entering, for example, in Eq.~(\ref{AvFluxProb1}), requires a ``change of variables'', from space-time location to flux space.
To this end, we exploit the fact that there is a straightforward relation between $\{t_S,\,{\bf x}_S\}$ and the
flux obtained by a single source, $\psi_S$, located at $\{t_S,\,{\bf x}_S\}$. For example,  Eq.~(\ref{AvFluxProb1}) can be rewritten in terms of the pdf for the flux from a single source, $p(\psi)$, as
\begin{equation}
\langle \Psi \rangle=\int_0^\infty d\Psi\, \Psi\, P(\Psi)\,=N\langle \psi\rangle =N \int_0^\infty d\psi\, \psi\, p(\psi)\,.\label{AvFluxProb2}
\end{equation}
Obtaining $p(\psi)$ (or rather its cumulative distribution) is the main subject of Sec.~\ref{sec:ppsi}. Let us briefly note  that the link is formally written as
\beq
p(\psi,\boldsymbol{x},t) =\frac{\nu}{N\,V_{\rm MW}}\int_{-t_{\rm MW}}^t d t_S\int_{\rm V_{\rm MW}} d{\bf x}_S \, \delta(\psi -q\;{\cal G}_B({\bf x}_S,t_S\to {\bf x},t))\,.
\eeq
 Once $p(\psi)$ is known, the pdf for the sum flux $P(\Psi)$ can be computed thanks to the convolution of the individual probabilities $p(\psi_i)$ under the constraint  $\sum_i^N \psi_i=\Psi$. The probability $P(\Psi)$ is formally written as
\begin{align}
P(\Psi)=\int_{\psi_1}\int_{\psi_2}\ldots
\int_{\psi_N}& p(\psi_1)\;p(\psi_2)\ldots \\&\ldots p(\psi_N)\;\delta\left(\sum_i^N \psi_i-\Psi\right)\;\mathrm{d}\psi_1\mathrm{d}\psi_2..\mathrm{d}\psi_N \;.\label{PNpsi}
\end{align}
This relation is based on the disputable yet natural assumption that the sources are not correlated with each other. The probability that two sources yield the fluxes $\psi_1$ and $\psi_2$ , respectively, is thus given by the product $p(\psi_1)\;p(\psi_2)$.

A second problem imposes itself as follows. If we try to use the generalization of Eq.~(\ref{AvFluxProb2}) to compute, for example, the second moment of the flux, $\langle \Psi^2 \rangle$,
and hence the expected variance of the flux, the expression formally diverges since the underlying $p(\psi)$  has a power-law tail,  $p(\psi)\propto\psi^{-\alpha - 1}$,  with $1<\alpha<2$ (as derived e.g., in \citepads{2012A&A...544A..92B}). It turns out that despite the fact that  (for instance) the variance of $p(\psi)$ is formally infinite, thanks to a so-called generalized central limit theorem,  the resulting probability $P(\Psi)$, for large $N$, has a universal shape, a so-called  { stable law}, only dependent on $\alpha$  and independent of $N$, but for a rescaling. More details on this are given in Sec.~\ref{sec:stablelaw}.
Hence, this difficulty does not appear so severe, since meaningful statistical quantities (such as confidence levels or quantiles) can still be computed in this limit. A relatively minor complication is that the index $\alpha$ of the pdf $p(\psi)$ is in fact energy dependent and depends on the propagation model as well. We introduce two limiting behaviors
of $\alpha$ in Sec.~\ref{sec:ppsi}, which allow for a satisfactory description of the distribution over a wide range of fluxes.

The third issue is that some of the above-mentioned properties depend on the fact that  the behavior $p(\psi)\propto\psi^{-\alpha - 1}$ formally extends to infinity. Physically, however, there is no such thing as an infinite flux: an obvious cutoff is imposed, for instance, by the empirically established absence of a source that is too close and/or too recent; of course, this depends on the level of credibility attributed to independent astrophysical information, such as available catalogs. In Sec.~\ref{sec:cut} and Sec.~\ref{sec:constraints}
we see the impact of this constraint on the maximum flux. A more subtle reason for having an effective cutoff to the domain of the probability distribution function $p(\psi)$  is the fact that Eq.~(\ref{DiffEq1}) is non-relativistic, and does not automatically ensure that causality is fulfilled. Accounting exactly for this is outside of the scope of our study, but we show that this introduces an effective cutoff that is particularly relevant at high energies (see Sec.s~\ref{sec:cut} and \ref{sec:causality}).
In either case, however, the conceptual problem that arises is that for a pdf $p(\psi),$ with a finite support, the asymptotic pdf for the sum of the fluxes variable $\Psi$ is now a Gaussian due to the standard central limit theorem (moments are, in fact, finite). It turns out that stable laws still provide acceptable descriptions of the pdf $P(\Psi)$ up to the fluxes of phenomenological interest, and that for the values of $N$ and $\Psi$ of actual interest one is relatively far from the Gaussian limit and much closer to the stable law limit. This aspect is described in Sec.~\ref{sec:cut}.

Before concluding this Section, a couple of comments are in order.

First, we shall limit ourselves to discussing the probabilities of departures of fluxes from their average only at a single energy. A natural generalization would be to discuss the probability that the flux at the Earth departs from its expected average value by more than a certain amount at two or more energies.
Fluxes at different energies (especially near ones) are not expected to be independent, rather strongly correlated, which can be formally  described, for instance, via the
following inequality in the conditional probability
\begin{align}
&P([\Psi(E_1)>\langle \Psi (E_1)\rangle] \cap[\Psi(E_2)>\langle\Psi (E_2)\rangle] ) \nonumber \\& \neq P(\Psi(E_1)>\langle \Psi (E_1)\rangle)\,P(\Psi(E_2)>\langle\Psi (E_2) \rangle) \:.
\end{align}
A manifestation of this property if second moments are finite is
\begin{equation}
\langle \Psi (E_1)\Psi (E_2) \rangle \neq \langle \Psi (E_1) \rangle \langle \Psi (E_2) \rangle \:.
\end{equation}
Estimating the probability of the most significant deviation among the different energy bins provides, however, an upper limit to the true probability of observing the actual flux excursion at different energy bins, the former being necessarily equal or lower due to the unitarity property of pdfs.

Second, while the effects discussed in our paper are the result of replacing the true underlying discrete source distribution with a continuum space-time distribution that describes its  smooth average, and are thus accounting for the partial (only ``probabilistic'') knowledge  we have on the sources frequency and position, this is different from what is usually intended by the ``effects of the granularity'' of the sources, in statistical mechanics, for example. These would typically manifest themselves in higher-order correlation functions (again, if finite, otherwise via conditional probabilities), like the correlations of fluxes measured at different times, or at the same time at different locations, etc. We emphasize that all source explosion models sharing the same time and volume average would give us the same flux $\langle\Psi\rangle$, while the above higher-order correlation observables could be different.
In particular, in the myriad model approach that we follow here, we assume point-like sources not to be correlated with each other. For example, a model where 20 SN explode at the same time in the Galaxy every 800 years does not satisfy this condition, and it will yield a different probability distribution function $P(\Psi)$ for the flux than a model where 1 SN explodes randomly in the Galaxy every 40 years. Both models have the same average flux but fluctuations around that mean do not follow the same law, and time correlations, for example, are different. On the other hand, our formalism can still provide a good approximation of strongly correlated source models: if the sources are strongly correlated in space and time (e.g., the 20 SN in the example above were to explode within a few pc distance of one another), simply downscaling $\nu$ and increasing $q$ by a corresponding factor would provide a satisfactory approximation for all but very small scale correlations.

\section{Technical aspects}\label{technical}

\subsection{Generalized central limit theorem: mathematical statement}\label{sec:stablelaw}
The central limit theorem states that the sum $\Psi$ of a large number $N$ of independent, identical, and stochastic variables $\psi$ is distributed according to a Gaussian law whose variance $\sigma_{\Psi}^{2}$ is $N$ times larger than the variance $\sigma_{\psi}^{2}$ of each individual contribution $\psi$. In our case though, the pdf of $\psi$ has a power-law tail $p(\psi) \propto \psi^{- \alpha - 1}$ with $1 < \alpha < 2$. The average flux $\langle \psi \rangle$ from a single source is defined but its variance is not and the central limit theorem cannot be applied, at least in its mundane form.
A more general form of the theorem has been discussed in section 1.8 of \cite{nolan2012stable}, for example, and can be readily applied to our problem. We simply recall here the salient features of the version that can be adapted to the myriad model.
Let $\psi$ be a random variable with probability law $p(\psi)$ defined on $\mathbf{R}_+$. The tail behavior can be captured by the cumulative distribution function (also called survival function) $C(\psi)$ such that
\beq
\forall \psi \geqslant 0, \;\; C(\psi) \equiv \int_{\psi}^{\infty} p(\psi') \; \mathrm{d}\psi' \;.
\label{eq:def_cumulative}
\eeq
That function encodes the probability for the flux of a single source to be larger than $\psi$. As shown in what follows, its asymptotic behavior, which turns out to be an essential ingredient in the proof of the generalized theorem, has the simple power-law form
\beq
\lim_{\psi \to \infty} \psi^{\alpha} C(\psi) = \eta >0 \; .
\label{eq:SL_definition}
\eeq
We denote by $\Psi = \psi_1 + \cdots + \psi_N$ the sum of $N$ independent and identically $p$ distributed random variables $\psi_i$. By introducing the rescaled flux
\beq
S_N=\frac{\Psi-\langle\Psi\rangle}{\sigma_N}, \quad\text{with}\quad \sigma_N=\displaystyle{\left(\frac{\eta\;\pi\; N}{2\Gamma (\alpha) \sin{(\alpha \, \pi/2)}}\right)^{1/\alpha}},
\label{eq:spread_GCLT}
\eeq
the generalized central limit theorem states that the probability law $P$ of $S_N$ converges for large $N$ toward the law $\mathbf{S}[\alpha,1,1,0;1]$ among the class of stable laws, in the notation of~\cite{nolan2012stable}. Note that the spread parameter $\sigma_N$ of the stable law depends on the constant $\eta$ and increases as $N^{1/\alpha}$ instead of $\sqrt{N}$ in the gaussian case. The general stable law function $\mathbf{S}[\alpha,\beta,\gamma,\delta;1]$ also depends on additional arguments beyond the stability index $\alpha$, notably a skewness parameter $\beta$, a scale parameter $\gamma$, and a location parameter $\delta$. The last index sets the type of parameterization used, as several exist in the literature. For instance, the normal or Gaussian distribution centered around $\mu$ and with variance $\sigma^2$ is denoted in this notation as $\mathbf{N}(\mu,\sigma)=\mathbf{S}(2,0,\sigma/\sqrt{2},\mu;1)$.
%
%
We recall that a pdf that is invariant under the product of convolution is described as stable. This is the case for well-known examples such as the Gaussian, Cauchy, and L\'evy distributions. But this class of functions is broad, as shown by Paul L\'evy in his study of sums of independent identically distributed terms in the 1920s \citep{levy1925calcul}. Most of the densities and distribution functions that it encompasses cannot be expressed analytically by closed-form expressions, although their Fourier transforms are tractable. That is why stable laws, such as the Pareto-Levy distribution $\mathbf{S}[\alpha,1,1,0;1],$ which we use hereafter, are not common in the field of astrophysics, but are relatively well known in finance for instance \citep{mandelbrot1960pareto,uchaikin1999chance}. Nowadays, computer programs such as Mathematica \citep{mathematica} can easily be used to compute these special functions.
%

\subsection{Application of the theorem to the probability of measuring a Galactic CR flux}\label{sec:ppsi}

\begin{figure}[h!]
\centering
\includegraphics[width=\columnwidth]{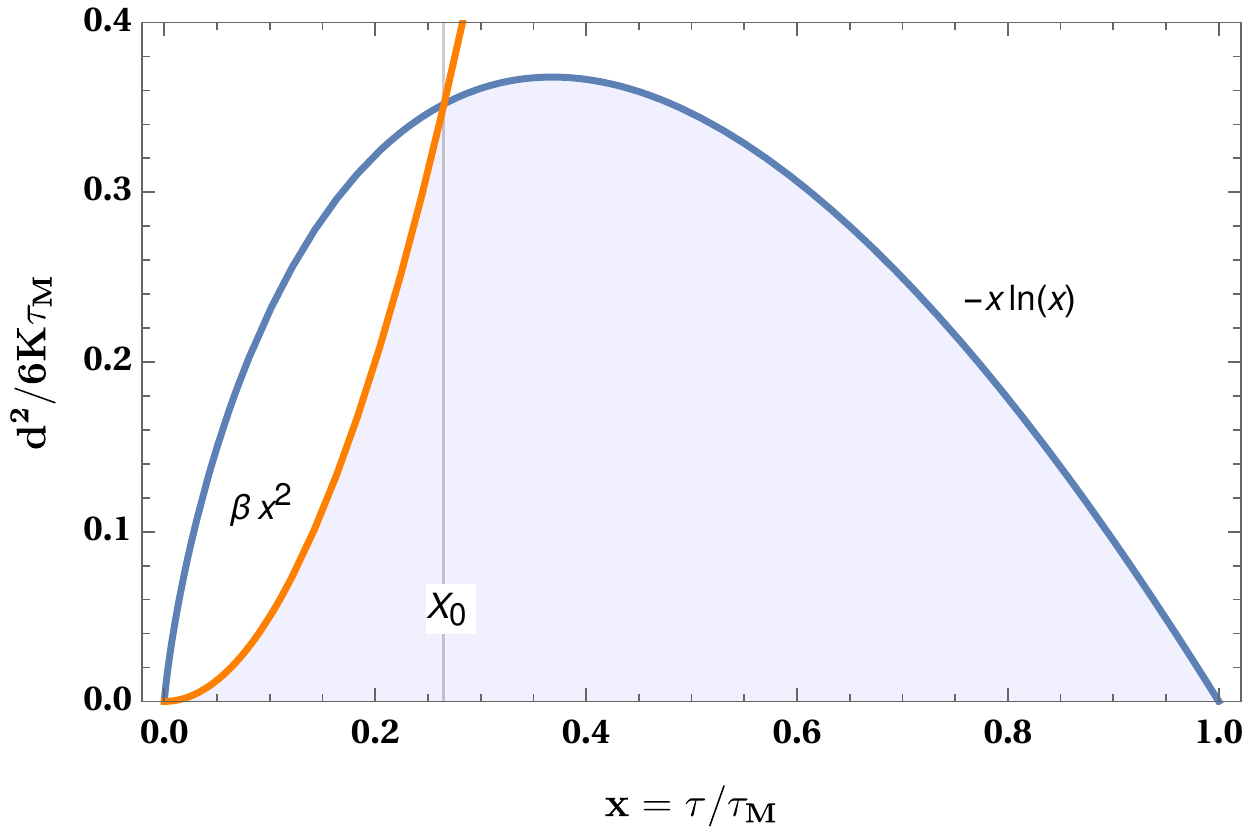}
\caption{Below the thick line: space and time region where a source contributes a flux $\psi' > \psi = {a}/{\tau_M^{3/2}}$ in the diffusion approximation.
Light blue shaded region: space and time domain that also respects the causal constraint.}
\label{fig:graf_cut_light_cone}
\end{figure}

To compute the spread $\sigma_N$ that comes into play in the stable law followed by the total flux $\Psi_N$, we need to determine the spectral index $\alpha$ so that $\lim\limits_{\psi \to \infty}C(\psi)\;\psi^{\alpha}=\eta$. We recall that $C(\psi)$ is the survival probability to get  a flux higher than $\psi  $ at the Earth. This probability can be obtained by integrating the density of a single source in space and time over the phase space region where it produces a flux larger than $\psi$.
As we are interested in the limit where the flux $\psi$ is large, if not infinite, the source must be local and young. In that case, CR propagation is the same as if the magnetic halo was infinite and diffusion was dominating the other processes such as convection and spallation.
The region in space and time that yields a flux $\psi$ may be derived from the boundless propagator of Eq.~(\ref{eq:prop3D}), which we translate into:
\beq
\psi=\frac{a}{\tau_M^{3/2}}\,x^{-3/2}\exp{\left(-\frac{d^2}{4 K \tau_M\,x}\right)}\;, \quad \text{with}\quad a=\frac{q}{(4\pi K)^{3/2}}\;.
\eeq
For simplicity, we have introduced the dimensionless time variable $x \equiv \tau/\tau_M$, where the timescale $\tau_M$ is defined as the maximal age for a source to provide a flux $\psi$. That value is reached when the source is located at $d=0$ and is equal to $\tau_M = (a/\psi)^{2/3}$.%
It is straightforward to check that for a given value $K$ of the diffusion coefficient, and hence at a given CR energy $E$, all space-time points in the plane $(x,d^2)$ satisfying the condition
\beq
\frac{d^2}{6\,K\,\tau_M}=-x\ln{x}
\label{eq:space_time_domain}
\eeq
are characterized by the same flux $\psi \equiv \psi(r=0,\tau=\tau_M)$.  This corresponds to  the thick blue line drawn in Fig.~\ref{fig:graf_cut_light_cone}. Points below that curve yield a flux $\psi'>\psi(0,\tau_M)$, while points above it produce a flux $\psi'<\psi(0,\tau_M)$.
One can thus compute the cumulative distribution $C(\psi)$ (sources with fluxes larger than $\psi$) by integrating the space and time density distribution of a single source $\rho{(\boldsymbol{r},t)}$ (assumed constant in the following) over the region in phase space located below the blue line of Fig.~\ref{fig:graf_cut_light_cone}. In the cylindrical approximation,  $N$ denoting the number of sources contributing to the total flux $\Psi$ allows us to write:
\beq
\rho{(\boldsymbol{r},t)}=\frac{\nu}{2 \, h \, \pi \, R^2}\frac{1}{N}\:.\eeq
There is no closed-form expression for $C(\psi)$ in general, but two relevant limiting situations can be discussed. Given $\psi$, if $\tau_M$ is such that the key length parameter $(6\,K\,\tau_M)^{1/2}$ is smaller than the typical thickness of the Galactic disk, $h$, the result will be equivalent to that of a uniform density in a 3D volume, hence
\begin{align}
C_{3D}(\psi) &= \int_\psi^{\infty} \! p(\psi') \, {\rm d}\psi' \nonumber\\&=
\left. \int_0^{\tau_M} \! {\rm d}t \, \frac{4}{3}\pi d^3 \, \rho{(\boldsymbol{r},t)} \, \right|_{\displaystyle \, d^2=-6K\tau_M x\ln{(x)}} \nonumber\\&=
\underbrace{\frac{1}{\pi^{1/3}}\;\frac{3^{3/2}\;}{2^{1/3}\;5^{5/2}}\;\frac{1}{N}\;\frac{\;R^{4/3}\;K^{2/3}}{h\;L^{5/3}\;\nu^{2/3}}\langle\Psi\rangle^{5/3}}_{\eta_{3D}}\;\psi^{-5/3}\;
\propto \psi^{-5/3}\:.
\label{eq:R_3D}
\end{align}
We highlight that the coefficient $\eta$ depends on the space-time probability density to find a source lying in the neighborhood of the observer; as does the typical flux spread $\sigma_N$. We may nevertheless express it as a function of the average flux $\langle\Psi\rangle$ to demonstrate how the various CR parameters at stake come into play. For the sake of clarity, the value of $\langle\Psi\rangle$ has been derived here assuming that the Galactic disk is an infinitesimally thick slab. Taking the 2D model neglecting radial boundary conditions leads to the simple expression
\beq
\langle\Psi\rangle=\frac{h\,L}{K}Q=\frac{q\,\nu\,L}{2\,\pi\,R^2\,K}\;,
\label{eq:slab_Psi}
\eeq
whose numerical value is very close to the actual result, as shown in the Appendix. Qualitatively, we expect that the behavior outlined in Eq.~(\ref{eq:R_3D}) is always attained for sufficiently high fluxes, since those require very close sources, although when this regime is attained depends on the energy $E$ via the diffusion coefficient $K$. Lower fluxes, however, are
also yielded by sources located ``far away'' from us, with respect to the scale $h$. In the limit where these dominate, one can effectively model the Galaxy as an infinitesimally thick disk, so that the only quantity that matters is the 2D (surface) density $\mu{(\boldsymbol{r},t)}$ which may be expressed as
\beq
\mu{(\boldsymbol{r},t)}=\frac{\nu}{\pi \, R^2}\frac{1}{N}\:.
\eeq
In this 2D limit, one gets:
\begin{align}
C_{2D}(\psi) &= \left. \int_0^{\tau_M} \! \mathrm{d}t \, \pi d^2 \, \mu{(\boldsymbol{r},t)} \, \right|_{\displaystyle \, d^2=-6K\tau_M x\ln{(x)}}\nonumber\\&=\underbrace{\frac{3\;}{2^{11/3}\pi^{2/3}}\;\frac{1}{N}\;\frac{K^{1/3}\;R^{2/3}}{\nu^{1/3}\;L^{4/3}}\;\langle\Psi\rangle^{4/3}}_{\eta_{2D}}\;\psi^{-4/3}\propto \psi^{-4/3}\:.
\label{eq:R_2D}
\end{align}
We argue in the following that, depending on the regime of fluxes $\Psi$ in which one is interested, either the 2D or the 3D distribution is relevant to the description of the problem. In any case, by virtue of the generalized version of the central limit theorem of Sec.~\ref{sec:stablelaw}, we know how to derive the pdf $P(\Psi)$ for both cases. It is sufficient to take an index $\alpha=5/3$ (3D) or  $\alpha=4/3$ (2D) and to compute the spread $\sigma_N$ from Eq.~(\ref{eq:spread_GCLT}) via the corresponding coefficients $\eta_{3D}$ and $\eta_{2D}$. Notice that $\sigma_N$ can be expressed as a function of the average flux $\langle \Psi \rangle$ and the various CR parameters. The number of sources $N$ has cancelled out in the product $\eta \, N$.
When it is large enough, the asymptotic regime where the central limit theorem holds is reached and we expect
\beq
P(\Psi)\to \;\frac{1}{\sigma_{N}}\;S[\alpha,1,1,0;1]\left(\frac{\Psi-\langle\Psi\rangle}{\sigma_{N}}\right)\;.
\label{eq:pPsi}
\eeq
In order to comment upon the dependencies of the spread $\sigma_N$ on the parameters of the problem, one may write
\beq
\frac{\sigma_N}{\langle \Psi \rangle} \; \propto \;
\begin{cases}
\;\displaystyle{\frac{K^{2/5}\;R^{4/5}}{\nu^{2/5}\;h^{3/5}\;L}}\, &\text{in the 3D case,}\\
\\
\;\displaystyle{\frac{K^{1/4}\;R^{1/2}}{\nu^{1/4}\;L}}\,  &\text{in the 2D case.}
\end{cases}
\eeq
As the reader may have noticed, the number $N$ of sources has disappeared from the rescaled quantity ${\sigma_N}/{\langle \Psi \rangle}$. The latter encodes the statistical excursions of the total flux $\Psi$ around its mean value $\langle \Psi \rangle$. Rescaled to the mean flux, the spread has the same dependence on the height $L$ of the diffusive halo ($\propto 1/L$) in both the 2D and 3D regimes. The thickness $h$ of the Galactic disk only enters in the expression of $\eta_{3D}$. The relative spread ${\sigma_N}/{\langle \Psi \rangle}$ increases with CR energy through the diffusion coefficient $K$. It decreases as the rate $\nu$ of explosions is increased. Both $K$ and $\nu$ enter through the ratio $K/\nu$, with an exponent of $2/5$ (3D) or $1/4$ (2D).

\subsection{The case with an upper cut on the flux}\label{sec:cut}

Strictly speaking, applying the generalized version of the central limit theorem requires that the cumulative distribution function $C(\psi)$ for a single source has a heavy-tail behavior up to an infinite flux $\psi$. This behavior should nevertheless break down since infinite fluxes, for example, are unphysical.
However, one expects the stable law expression~(\ref{eq:pPsi}) for $P(\Psi)$ to still provide a good approximation of the actual distribution up to some value of the total flux $\Psi$, should the underlying power-law behavior of the single source pdf $p(\psi)$ be valid over a sufficiently large range. To quantify this, let us assume that the single source pdf follows the power-law behavior $p(\psi) \propto \psi^{-\alpha - 1}$,  with $1<\alpha<2$, up to some critical value $\psi_{cut}$ above which it vanishes. There may be several reasons for this heuristic argument to hold, some to be explored in the following. Schematically, one can write the probability of measuring a flux $\psi$ from a single source as:
\beq
p(\psi)=(1+\epsilon)\;p_{th}(\psi)\;\Theta(\psi_{cut}-\psi)\:,\label{ansatz1}
\eeq
where $p_{th}(\psi)$ is the idealized pdf with an infinite power-law tail, discussed in section~\ref{sec:stablelaw}. Of course this may be too crude an approximation close to $\psi_{cut}$, but provided that $\psi_{cut}$ is large compared to the fluxes of interest and the renormalization correction $\epsilon$ is small, it should not have a significant affect. In fact, given the Ansatz of Eq.~(\ref{ansatz1}), one can write the probability to measure a flux $\Psi=\sum_i^N \psi_i$ from N sources, among which ${N}$ are located inside the light cone, as
\begin{align}
P(\Psi)&=\int_{\psi_1}\ldots \int_{\psi_{N}}p(\psi_1) \ldots p(\psi_{N})\;
\delta\left(\sum_i^{N} \psi_i-\Psi\right)\;\mathrm{d}\psi_1 \ldots\mathrm{d}\psi_{N}\;,
\label{PDF_Psi_yoann_1}
\;\\
&=(1+\epsilon)^{{N}}\;\int_{\psi_1}\ldots  \int_{\psi_{N}}p_{th}(\psi_1)\ldots p_{th}(\psi_{N})\;
\Theta(\psi_{cut}-\psi_1)\ldots \nonumber\\ &\ldots\Theta(\psi_{cut}-\psi_{N})\;\delta\left(\sum_i^{N} \psi_i-\Psi\right)\;\mathrm{d}\psi_1\ldots \mathrm{d}\psi_{N} \:.
\label{PDF_Psi_yoann_2}
\end{align}
For values of $\Psi$ such as $\Psi<\psi_{cut}$\footnote{This condition is denoted hereafter by a $(\star)$}, each individual value $\psi_i$ must satisfy $\psi_i<\psi_{cut}$. Hence we get

\begin{align}
P(\Psi)=&(1+\epsilon)^{{N}}\;\int_{\psi_1}\ldots \int_{\psi_{N}}p_{th}(\psi_1)\nonumber\\&\ldots p_{th}(\psi_{N})\;\delta\left(\sum_i^{N} \psi_i-\Psi\right)\;\mathrm{d}\psi_1\ldots \mathrm{d}\psi_{N} \:\:{\rm for}\:\:\Psi<\psi_{cut}\;.
\end{align}

\noindent
Thus, for { sufficiently} large $N$ , we expect
\begin{align}
P(\Psi)&\approx(1+\epsilon)^{{N}}\;\frac{1}{\sigma_{N}}\;S[\alpha,1,1,0;1]\left(\frac{\Psi-\langle\Psi\rangle_{th}}{\sigma_{N}}\right)\nonumber\\ &\quad\text{for}\quad\Psi<\psi_{cut}\;(\star)\:,
\label{eq:main_relation}
\end{align}
where the average flux $\langle\Psi\rangle_{th}$ in the argument of the stable function corresponds to the uncut idealized pdf $p_{th}(\psi)$.
The approximation of Eq.~(\ref{eq:main_relation}) is relatively general, but its actual usefulness depends on the precise values of $\psi_{cut}$ and $\epsilon$. For example, as long as $N \epsilon \ll 1$, the result does not really depend on $\epsilon$. That condition turns out to be satisfied for the cases that are subsequently discussed. In the following subsections, we actually distinguish two evident physical reasons for imposing a cut on the tail of the pdf distribution.

\subsection{Causality effect}\label{sec:causality}

To commence, propagation through diffusion should not violate causality, a condition which is not always satisfied by the propagator as defined in Eq.~(\ref{eq:prop3D}). Taking any finite age $\tau$ for a source leads to a non-zero value for ${\cal G}$, whatever the distance $d$. This will not cause any problems for far and young sources for which the flux is exponentially suppressed, but  for young and close objects, which happen to dominate large positive fluctuations of $\Psi$ with respect to the average value $\langle\Psi\rangle$, this can lead to much larger flux contributions than
physically allowed.

A correction for this effect can  be quantified with reference to the space-time diagram of Fig.~\ref{fig:graf_cut_light_cone} by removing the portion where causality is violated from the region
of phase space that contributes to the flux. The domain to be withdrawn extends above the orange curve defined by the light cone condition $d^2=c^2 x^2 \tau_M^2$, $c$ being the speed of light in this case. The orange parabola allows us to carve away the domain in white that lies below the blue line, leaving only the light-blue shaded region to contribute to fluxes larger than $\psi$. The intersection between the blue (diffusion) and orange (light cone) curves takes place at $x = x_{0}$ for which
\beq
-\beta\,x_0= \ln{(x_0)} \quad\text{     with  }\quad\beta=\frac{c^2\,\tau_M}{6K}\:.
\label{eq:cond_x0}
\eeq
The cumulative distribution function $C(\psi)$ has to be recalculated. It is given now by two different contributions corresponding to the integrals of the light-blue shaded area extending to the left ($C_1$) and to the right ($C_2$) of the vertical line at $x_0$ in Fig.~\ref{fig:graf_cut_light_cone}. The 2D and 3D results may be expressed as:
\begin{align}
C_{2D}^{\text causal}(\psi)=&\;\pi\,\mu_{(\boldsymbol{r},t)}\; 6K\;\left(\frac{a}{\psi}\right)^{4/3} \;\left\{ \left(\frac{1-x_0^2}{4}\right)\;+\;\frac{x_0^2}{2}\,\ln{x_0}\right\}\;\nonumber \\&+\;\pi\,\mu_{(\boldsymbol{r},t)}\;c^2\;\left(\frac{a}{\psi}\right)^{2}\frac{x_0^3}{3} \equiv C_{1}^{2D}(\psi)\,+\,C_{2}^{2D}(\psi)\;,
\end{align}

\begin{align}
C_{3D}^{\text causal}(\psi)=&\frac{4}{3}\,\pi\,\rho_{(\boldsymbol{r},t)}\; (6K)^{3/2}\;\left(\frac{a}{\psi}\right)^{5/3}  \;\int_{x_0}^{1}\{-x\ln{(x)}\}^{3/2}\mathrm{d}x\;\nonumber \\&+\;\frac{4}{3}\,\pi\,\rho_{(\boldsymbol{r},t)}\;c^3\;\left(\frac{a}{\psi}\right)^{8/3}\frac{x_0^4}{4}\equiv C_1^{3D}(\psi)\,+\,C_2^{3D}(\psi)\;.
\end{align}

The larger the flux $\psi$, the smaller the maximal age $\tau_M$. In this limit, the coefficient $\beta \propto \tau_M$ vanishes. As the orange parabola of Fig.~\ref{fig:graf_cut_light_cone} opens, the value of $x_0$ tends to 1 and a large fraction of the space-time volume becomes causally disconnected.  Thus, in the high flux limit, $C_1$ vanishes while $C_2$ increases. In the 2D regime, $C_{2D}(\psi)\propto\psi^{-2}$ whereas in the 3D regime, $C_{3D}(\psi)\propto\psi^{-8/3}$. As mentioned previously, the latter eventually takes over the former for very large values of the flux $\psi$ for which the typical distances of the sources are smaller than the thickness $h$ of the Galactic disk. As $8/3>2$, the variance associated to the pdf $p(\psi)$ is now finite.
According to the central limit theorem, the probability $P(\Psi)$ converges toward a Gaussian when the number of sources $N$ goes to infinity. We note, incidentally, that if the Galactic disk was infinitesimally thick, with $h=0$, the 2D regime would apply with $\alpha=2$. The variance is in that case divergent but the generalized central limit theorem can be applied, with the consequence that the total flux pdf $P(\Psi)$ also reaches a Gaussian form.
Although causality arguments {\it per se}  would allow arbitrarily large fluxes, we see that the whole discussion on stable laws loses its importance once a sizable fraction of the space-time volume is removed, since $p(\psi)$ is too steep. In practice, one can account for these effects by abruptly cutting the ``standard'' power-law distribution pdf above a transition flux $\psi_{cut}\equiv\psi_c$, which we may define for instance via the condition $C_1(\psi_c)=C_2(\psi_c)$. As we are interested in computing probabilities around the mean value $\langle\Psi\rangle$, let us  compute $\psi_c$ and compare it with $\langle\Psi\rangle$.
To do so, we need to determine ($\psi_c, x_c$) by solving the two following equations:
\beq
C_1(\psi_c)=C_2(\psi_c)\text{  and  }-\beta_c(\psi_c)\,x_c= \ln{(x_c)}\:,
\eeq
which numerically yields
$x_c\approx0.6226$ in the 2D case and
$x_c\approx0.6424$ in the 3D case.
Note that $x_c$ (or equivalently $\beta_c$) does not depend on the cosmic ray energy, and is not very different between the 2D and 3D regimes.
We can also define $y=\psi_c/\langle\Psi\rangle$, which can be written as
\beq
y=\;\frac{\psi_c}{\langle\Psi\rangle}\;=\;\frac{1}{4\,6^{3/2}\!\sqrt{\pi }}\;\frac{1}{\beta_c^{3/2}}\;\frac{c^3\,R^2 }{K^2\,L\,\nu  }\;.
\eeq
The estimate for $y$ varies only by 14\% between the 2D and 3D cases. As this difference is relatively small, we will take the average between these two values in the following discussion. Notice that $y$ depends on energy through the diffusion coefficient $K$.
Of course, for our discussion to be of any relevance, the stable law Eq.~(\ref{eq:main_relation}) should be valid for a total flux $\Psi$ well in excess of its average $\langle\Psi\rangle$, that is, for values of $y$ as large as possible. As an example, for the MED propagation model borrowed from \citeads{2004PhRvD..69f3501D}, we find that $y\simeq32$ at 10 TeV while it reaches $\sim 2 \times 10^{4}$ at 100~GeV.

\begin{figure}[h!]
\centering
\includegraphics[width=\columnwidth]{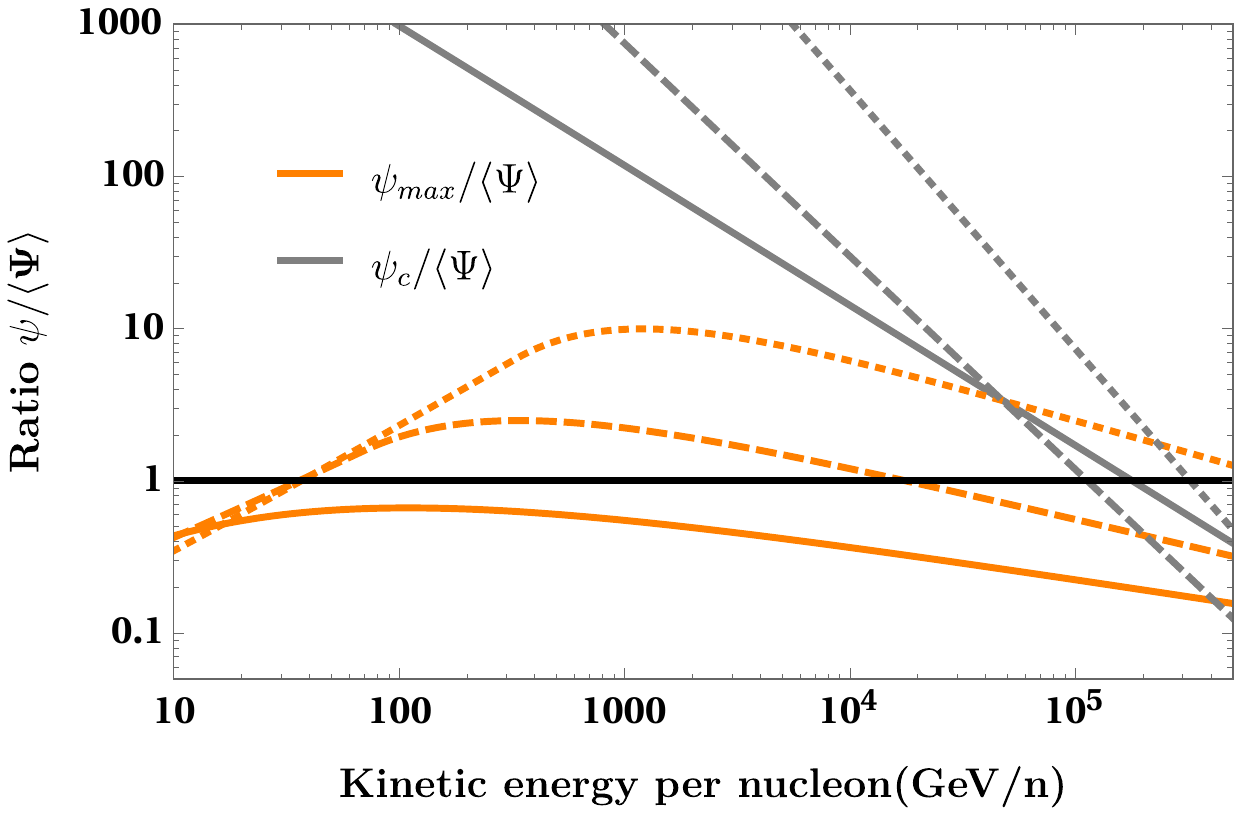}
\caption{
The causal cutoff flux $\psi_c/\langle\Psi\rangle$ (local environment cutoff flux $\psi_{max}/\langle\Psi\rangle$) as a function of energy is displayed in gray (orange) from top to bottom for the three propagation models MIN (dotted), MED (dashed) and MAX (solid).
}
\label{fig:ratio_psi_c_psi_env}
\end{figure}

The evolution of $\psi_c/\langle\Psi\rangle$ as a function of energy is displayed in gray in Fig.~\ref{fig:ratio_psi_c_psi_env} for the three different benchmark propagation models discussed in \citeads{2004PhRvD..69f3501D}, namely MIN, MED and MAX, and whose parameters are in Tab.~\ref{tab:model:annexe} of the Appendix. That ratio is always larger than ten for cosmic ray energies below 10~TeV. To be conservative, the causality constraint could be forgotten below the TeV scale as long as we are interested in values of $\Psi$ not exceeding a few times the average. However, depending on the propagation model, the light cone cut-off may seriously impair the use of Eq.~(\ref{eq:main_relation}) above a few tens of TeV, an energy range probed by calorimetric instruments such as CREAM or CALET.

\subsection{Absence of very close and/or young sources}\label{sec:constraints}
Another natural limitation to the maximum flux that a single source may contribute to the measured CR flux $\Psi$ comes from some (partial) deterministic information on nearby sources; {for example, one knows the sources within some distance one by one, but not the ones that are far away; or one only knows that no source exists within some distance}. In a schematic way, one may split the flux into the sum of a local contribution from known sources and a remote component, on which the only information is of statistical nature, as considered above:
\beq
\Psi=\Psi_{\rm loc}+\Psi_{\rm far}\;.
\eeq
The situation may be intermediate, for instance one has a catalog whose completeness decreases with distance and age of the source, but accounting for this complication is unnecessary for what follows. For a given catalog, which is assumed to be complete within a given region of space-time, from which  the contribution $\Psi_{\rm loc}$ can be ``exactly'' computed, one could repeat the reasoning of the previous section to determine the pdf associated to $\Psi_{\rm far}$. The cumulative distribution function $C(\psi)$ of a remote source can be computed by carving out the space-time region covered by the catalog.

\begin{figure}[h!]
\centering
\includegraphics[width=\columnwidth]{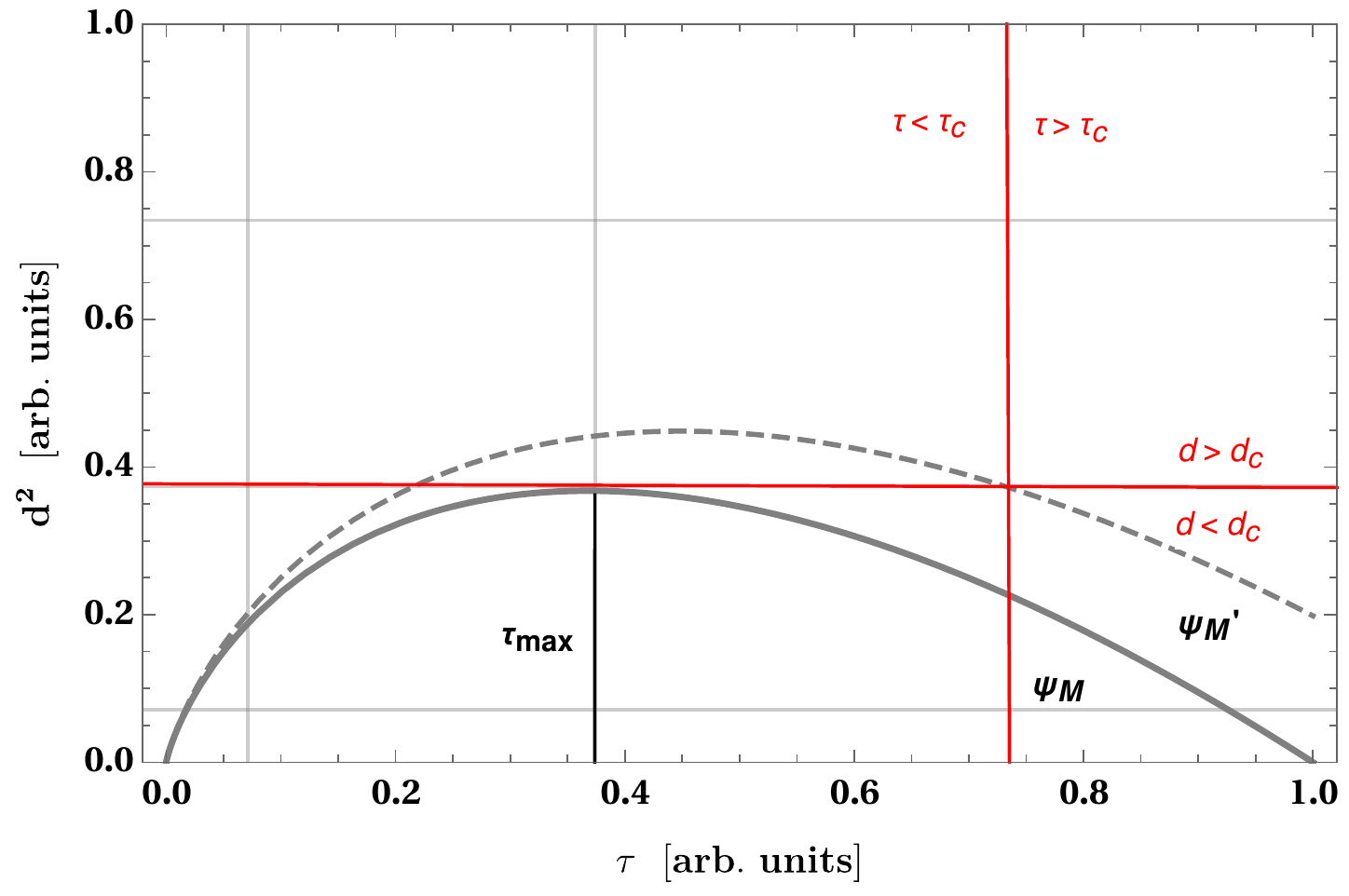}
\caption{ Solid gray line: locus of source ages and distances giving the same flux as the maximal one $\psi_M$ yielded by a source located at distance $d_c$ from the Earth, attained for $\tau=\tau_{max}={d_c^2}/{6K}$. Dashed gray line: locus of source ages and distances giving the same flux $\psi_M'$ as the one of a source of age $\tau_c$ and distance $d_c$. In the case shown ($\tau_{max}<\tau_c$), the flux $\psi_M'$ is also the maximal flux from a source farther than  $d_c$ and older than $\tau_c$.
}
\label{fig:max_psi}
\end{figure}

Let us develop instead a slightly simpler argument by determining the maximal flux $\psi_{cut}\equiv\psi_{max}$ one can expect from a source located at the inner boundary of the phase space region filled by the catalog. We can use it to find the closest and youngest object, hence deriving plausible values for the minimal distance and age below which no cosmic ray injection takes place. A source located at distance $d,$  which exploded at time $\tau$ in the past contributes the flux
\beq
\psi\;=\;\frac{q}{(4\,\pi\,K\,\tau)^{3/2}}\exp{\left(-\frac{d^2}{4 K \tau} \right)}\;.
\eeq
In the most general case, we want to put lower limits on the distance ($d>d_c$) and age ($\tau>\tau_c$) of remote sources in order to extract the cut-off, hereafter denoted by $\psi_{max}$, above which the pdf $p(\psi)$ vanishes. Constraining the distance $d$ to be larger than $d_c$ translates into the maximal flux
\beq
\psi_{M}={\displaystyle \frac{q}{({2\,\pi\,d_c^2}/{3})^{3/2}}} \exp{\left(-\frac{3}{2} \right)}\;,
\label{eq:psi_M}
\eeq
which is reached for an age $\tau_{max}={d_c^2}/{6K}$. The flux $\psi_{M}$ corresponds to the solid gray curve of Fig.~\ref{fig:max_psi}. In this space-time diagram, the region below $d = d_c$ is excluded. From now on, two cases must be distinguished depending on the relative values of $\tau_{max}$ and $\tau_{c}$.
If $\tau_{max}>\tau_c$, the maximum flux that a remote source may provide is still $\psi_{M}$, a value reached at the space-time location ($\tau_{max},d_c$).
On the other hand, if $\tau_{max}<\tau_c$, the region to the left of the vertical line $\tau=\tau_c$ being excluded, remote sources can no longer yield the flux $\psi_{M}$. The solid gray line lies entirely in the excluded portion of the phase space diagram of Fig.~\ref{fig:max_psi}. In this regime, the maximal flux attainable by a remote source is $\psi_{M}'<\psi_{M}$. It corresponds to the dashed gray line that touches the allowed phase space region at location ($\tau_c,d_c$).

\noindent
In summary, imposing the constraint $\{d>d_c,\: \tau>\tau_c\}$ translates into $\psi<\psi_{max}$ with:
\begin{align}
\psi_{max}=\begin{cases}
\psi_{M} \equiv \displaystyle\frac{q}{\left(2\,\pi\,d_c^2 \right/3)^{3/2}}\exp{\left(-\frac{3}{2} \right)}\;,
&\text{  if } \underbrace{d_c^2/{6K}}_{\tau_{max}}>\tau_c\:,\\
&\\
\psi_{M}' \equiv \displaystyle\frac{q}{(4\,\pi\,K\,\tau_c)^{3/2}}\exp{\left(-\frac{d_c^2}{4\,K\,\tau_c} \right)}\;,
&\text{  if } \tau_{max}<\tau_c\:.
\end{cases}
\label{eq:max_psi_1}
\end{align}
This value may be compared to the mean flux $\langle\Psi\rangle$ taken as the theoretical average corresponding to the slab model and derived in Appendix~\ref{annexe:1}. The relations~(\ref{eq:max_psi_1}) readily translate into
\begin{align}
{\displaystyle \frac{\psi_{max}}{\langle\Psi\rangle}}=\begin{cases}
{\displaystyle \frac{3\,R^2\,K}{\nu\,L\,d_c^3}} \sqrt{{\displaystyle \frac{3}{2\,\pi}}} \exp{\left(-{\displaystyle \frac{3}{2}} \right)}\;,
&\text{  if } \tau_{max}>\tau_c\;,\\
&\\
{\displaystyle \frac{R^2}{4\,\nu\,L\,\sqrt{\pi\,K\,\tau_c^{3}}}}
\exp{\left(-{\displaystyle \frac{d_c^2}{4\,K\,\tau_c}}\right)}\;,
&\text{  if } \tau_{max}<\tau_c\;.
\end{cases}
\label{eq:max_psi_2}
\end{align}

As a conservative example of this procedure, we have taken $d_c=0.06\,{\rm kpc}$ as our lower limit on the distance, inspired by the closest known source G+276.5+19.  In the same way, our lower boundary on the age comes from the youngest known source J0855-4644 for which $\tau_c=2.7\,{\rm kyr}$. These values are taken from the catalog compiled by~\citet{2010A&A...524A..51D}. Using relations~(\ref{eq:max_psi_2}), we have derived the ratio $\psi_{max}/\langle\Psi\rangle$ and plotted it in orange in Fig.~\ref{fig:ratio_psi_c_psi_env} to compare it with $\psi_{c}/\langle\Psi\rangle$.
We observe a change of slope in the dependence of $\psi_{max}/\langle\Psi\rangle$ as a function of cosmic ray energy. This is particularly obvious in the MIN case (dotted orange curve) for which the diffusion coefficient is the smallest at low energies compared to the other propagation regimes MED and MAX. In the MIN configuration, the ratio $\psi_{max}/\langle\Psi\rangle$ increases with energy below 1~TeV. In this regime, $K$ is small and the critical age $\tau_{max}={d_c^2}/{6K}$ exceeds the lower bound $\tau_c$. According to Eq.~(\ref{eq:max_psi_2}), the ratio $\psi_{max}/\langle\Psi\rangle$ scales as $K$ and increases with cosmic ray energy. At approximately 1~TeV, a change of regime occurs when $\tau_{max}$ becomes smaller than $\tau_c$, and the ratio $\psi_{max}/\langle\Psi\rangle$ scales as $1/\sqrt{K}$, decreasing with energy. The same trend is featured by the MED (dashed orange) and MAX (solid orange) curves, although in a milder way.

As clearly shown in Fig.~\ref{fig:ratio_psi_c_psi_env}, the cut-off $\psi_{max}$ imposed by the catalog of local and recent sources is much more constraining than the value $\psi_c$ yielded by the causality argument. The orange curves are basically always below the gray lines, which they intersect at energies in excess of 200~TeV, and for values of the ratio $\psi_{cut}/\langle\Psi\rangle$ of approximately 1.
The use of approximation~(\ref{eq:main_relation}) is therefore possible below the orange curves. It is effectively interesting for values of the cut-off $\psi_{max}$ larger than the average flux $\langle\Psi\rangle$. This is not the case for the MAX configuration. For the MIN and MED propagation models, the stable law~(\ref{eq:main_relation}) can still be applied on a fairly limited region in energy and total flux $\Psi$. As featured in Fig.~\ref{fig:ratio_psi_c_psi_env}, the ratio $\psi_{max}/\langle\Psi\rangle$ reaches a maximal value of ten at an energy of 1~TeV. The constraint from the catalog is therefore relatively strict. Most of the populations of sources that would otherwise lead to large values of the cosmic ray flux are excluded from the statistical analysis when the age and distance constraints are imposed.
A word of caution is in order though since the completeness of a catalog is always questionable. Setting lower limits on the age and distance of nearby sources may be relatively subjective and eventually hazardous. That is why we have disregarded them in the Monte Carlo simulations that we discuss in the following section.

\section{Comparison with numerical simulations}\label{simulations}

Until now, we have determined the range of fluxes $\Psi$  for a given propagation setup over which the approximation described in Sect.~\ref{sec:cut} holds. However, identically to how the combined pdf $P(\Psi)$ tends toward a Gaussian law when the conventional central limit theorem applies, the convergence to a stable law is only an asymptotic behavior. While exact results on the closeness to a stable law at finite $N$ may exist in the mathematical literature, generic results are not useful in our case where a cut-off may be imposed on the individual flux pdf $p(\psi)$.
We thus need to validate the reliability of the analytical theory via extensive numerical simulations. Our aim is to study how the total flux pdf $P(\Psi)$ converges to or departs from the stable law Eq.~(\ref{eq:pPsi}).

\begin{figure*}[h!]
\centering
\includegraphics[width=1.95\columnwidth]{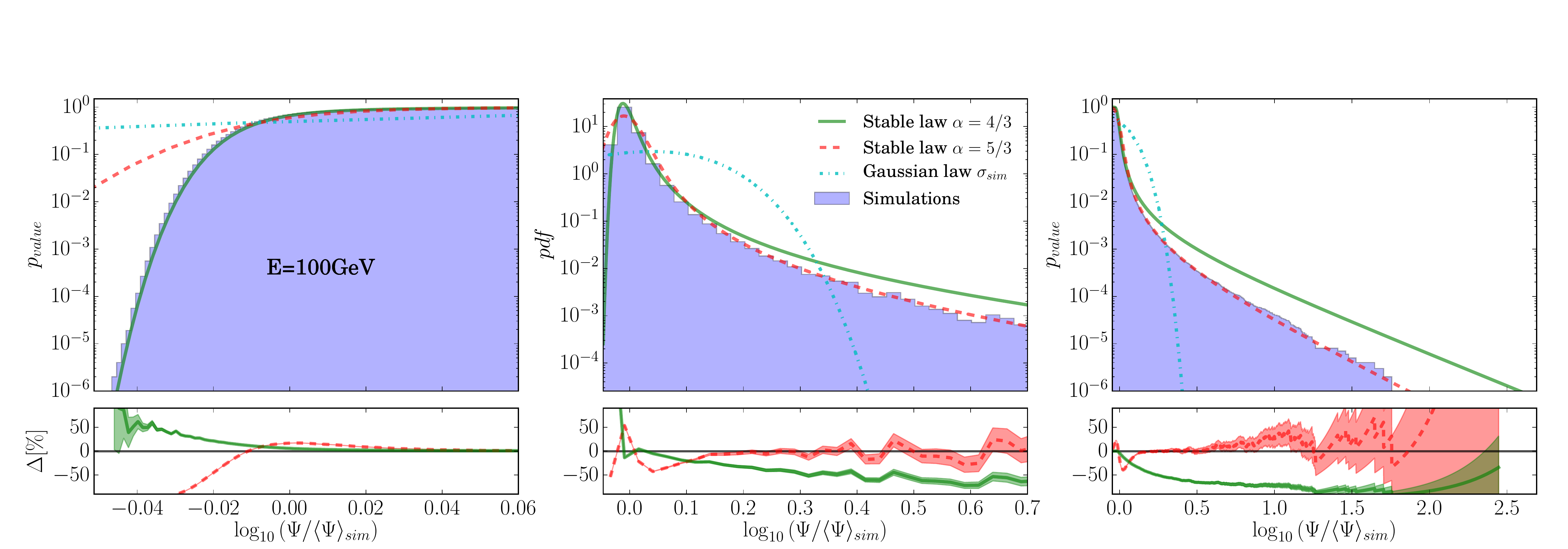}
\includegraphics[width=1.95\columnwidth]{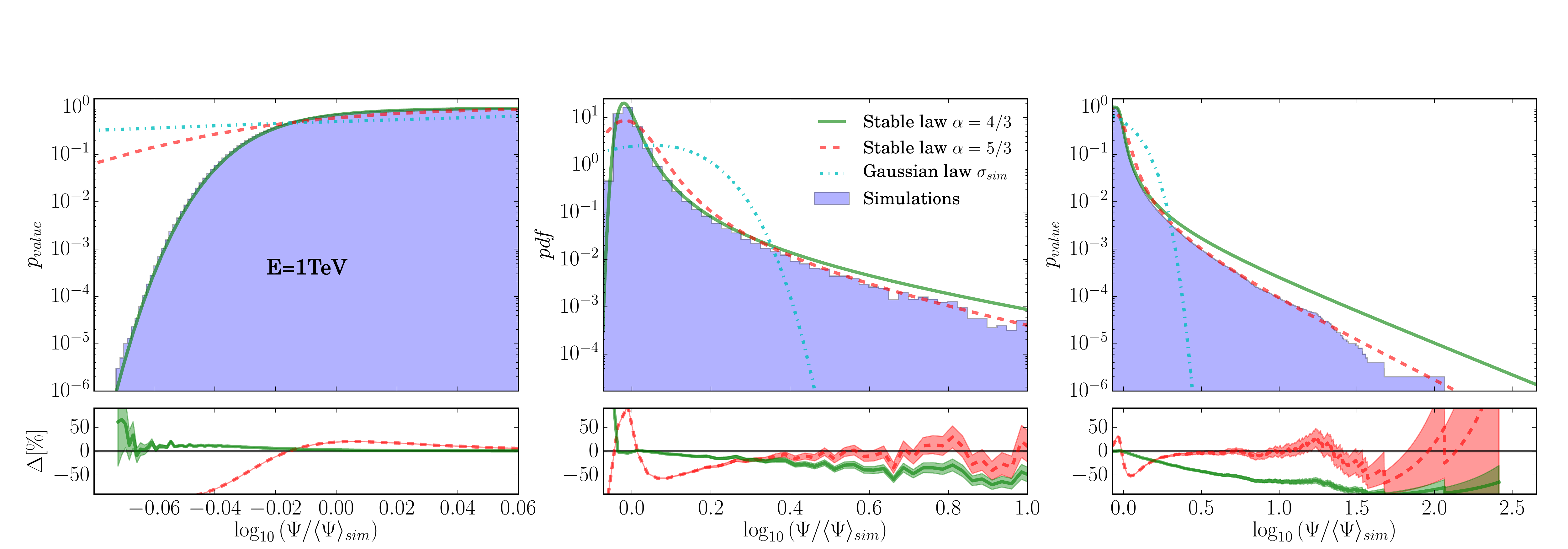}
\includegraphics[width=1.95\columnwidth]{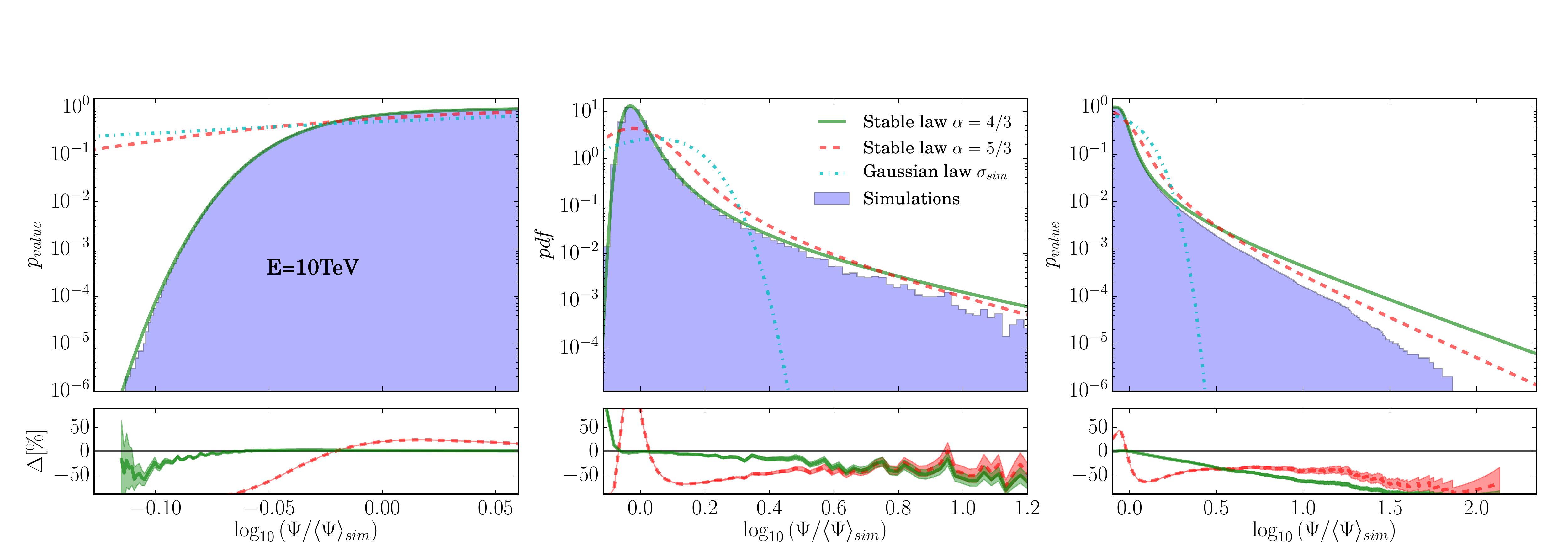}
\caption{For each row, the left and right cumulative blue histograms of $10^6$ Monte Carlo realizations of Galactic populations of CR sources are displayed in the left and right panels, respectively, whereas the pdf $P(\Psi)$ stands in the middle. The MED propagation model is used without taking into account convection, diffusive reacceleration and spallations. From top to bottom, the CR kinetic energy has been set equal to 100~GeV, 1~TeV, and 10~TeV. The solid green line indicates the theoretical prediction for the 2D model of the Galactic magnetic halo, whereas the dashed red curve corresponds to the 3D case. The residuals between theory and simulations are displayed below each histogram with their 1-$\sigma$ Poissonian error.}
\label{fig:simu_causal_t0_a}
\end{figure*}
\begin{figure}[h!]
\centering
\includegraphics[width=\columnwidth]{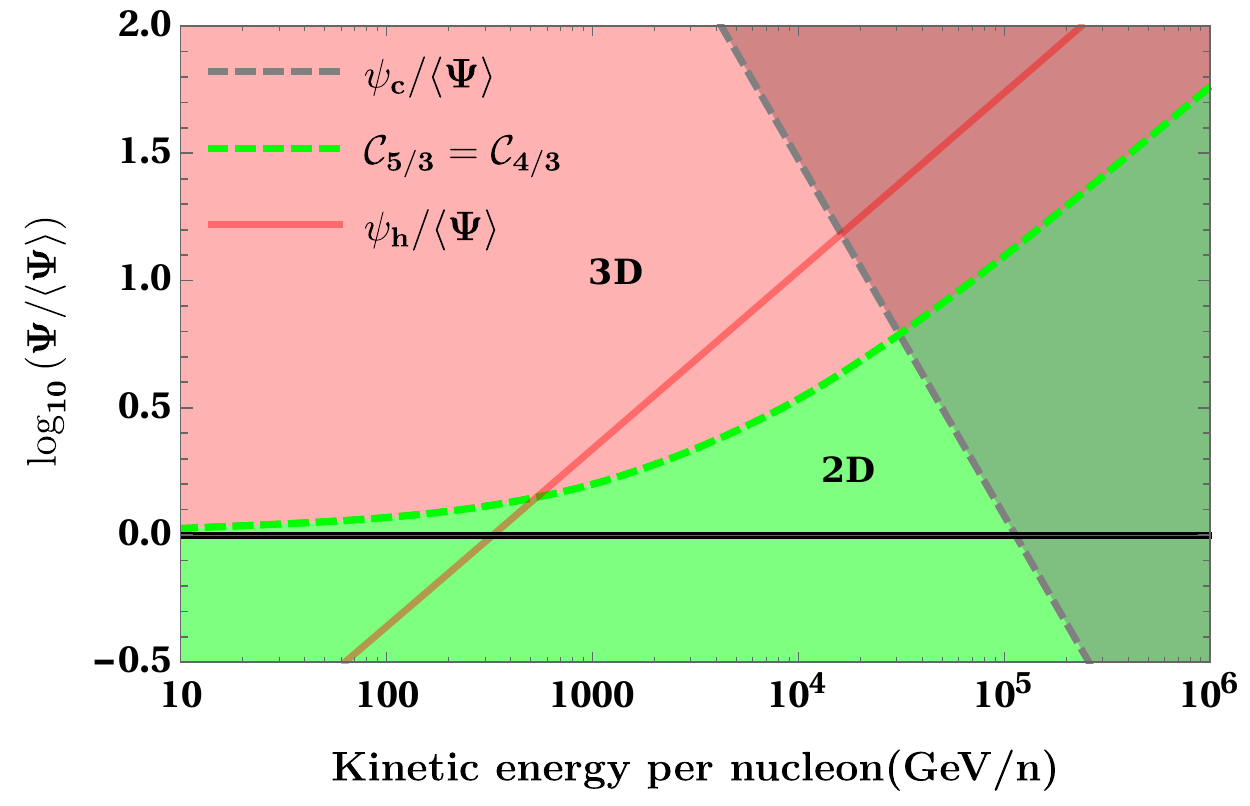}
\caption{
Regions in the flux versus energy plane where the 2D or 3D stable law is best suited to computing the probability of an excess above the mean (MED propagation model assumed). For a fluctuation lying in the light red (green) region, one should use the 3D (2D) approximation corresponding to the index $\alpha=5/3$ ($\alpha=4/3$). The dashed green curve signals the transition between these regimes as estimated from the equality of the cumulative distributions, ${\cal C}_{5/3}={\cal C}_{4/3}$. The solid red line reports the alternative estimate of $\psi_h$ following the argument developed in the text.
In the shaded area in the upper-right corner of the diagram, causality is expected to generate deviations from the stable law behavior as a result of the light cone cut-off $\psi_c$ it implies on the pdf.}
\label{fig:Transition_2d_3d_psic}
\end{figure}
\subsection{Simulation settings}
For each realization of $\Psi$, we simulate a Galactic population of sources in the framework of the 3D model of Eq.~(\ref{eq:source_distribution}). Each source is generated at the random position $\bf x_i$ within the Galactic disk, and with the random age $\tau_i$. The positions $\bf x_i$ are taken within an idealized cylindrical Milky Way Galaxy of radius $R$ and half-thickness $h$, so that the spatial density of sources is homogeneous within the disk. The age of each source $\tau_i$ is taken between 0 and $T$, where $T$ is the integration time of the simulation and is chosen to be ${3\;\pi L^{2}}/{4\;K}$, that is, approximately three times the typical diffusion time within the magnetic halo. The number of sources of a realization is $N=T\nu$, with $\nu$ the rate of SN explosions, taken here to be three per century. Note that for a constant explosion rate of 3 SN/century and the choosen $T$, this means considering only sources younger than about 133 Myr at 100 GeV, so still much younger than the age of the Galaxy. In our model, this number reaches approximately $4\times 10^6$ at a reference energy of 100 GeV. The spectrum of particles injected by each source is defined as
\beq
q = q_0 \, \left( {\displaystyle \frac{\cal R}{\rm 1 \, GV}} \right)^{-2.2} \;,
\label{eq:q_injection}
\eeq
where ${\cal R}$ stands for the CR rigidity while the normalization $q_0$ is set equal to 1 GeV$^{-1}$. Notice that $q$ does not come into play insofar as it can be factored out from our numerical results. The flux yielded by a source is computed assuming the CR propagation model MED, where we have safely neglected the effects of convection, diffusive reacceleration and spallations, which become negligible at the energies considered in our simulations. Without loss of generality, fluxes are derived for an observer placed at the radial and vertical centers of the Galactic disk, that is, at $r = z = 0$. The contributions $\psi_i$ of the $N$ sources of a given population are added to obtain one realization of $\Psi$. When the causality condition discussed in Sec.~\ref{sec:causality} is implemented, the contributions from sources lying outside the light cone of the observer are not included. We evaluated $\Psi$ for each decade of energy between 100 GeV and 1 PeV.
As we are interested in the pdf of the random variable $\Psi$, we simulated ``only'' $10^6$ realizations of an idealized cylindrical Milky Way Galaxy, since we are not concerned with probabilities below the $10^{-5}$ level.

\subsection{Simulation results}

In Fig.~\ref{fig:simu_causal_t0_a}, we show the results of the simulations that we carried out at 100 GeV, 1 TeV, and 10 TeV, with the light cone cut-off switched on. In each row, the left and right cumulative blue histograms of $10^6$ realizations of Galactic populations of CR sources are displayed in the left and right panels, respectively, whereas the pdf $P(\Psi)$ stands in the middle. Once normalized to unity, the cumulative histograms on the left and on the right directly yield the probability of getting a population sourcing a flux smaller or larger,  respectively, than the value $\Psi/\langle\Psi\rangle_{sim}$ read on the horizontal axis. Notice that $\langle\Psi\rangle_{sim}$ denotes the value of the flux averaged over the $10^{6}$ realizations of our Monte Carlo. Due to finite sampling, results start to become unreliable for probabilities below $10^{-5}$, and are obviously not even defined below $10^{-6}$.

In each of the panels of Fig.~\ref{fig:simu_causal_t0_a}, we also display the left and right cumulative distributions as well as the pdf corresponding to the theoretical predictions of Sec.~\ref{sec:ppsi}. The dashed red line stands for the 3D case (i.e., $\alpha=5/3$) whereas the solid green curve refers to the 2D case (i.e., $\alpha=4/3$). In the argument of the stable distribution of Eq.~(\ref{eq:pPsi}), the average flux $\langle\Psi\rangle$ must be calculated exactly, assuming the same Milky Way magnetic halo as in the Monte Carlo simulations, namely a disk of radius 20~kpc and a maximal age for the SN explosions of $T$. Depending on whether the theoretical prediction is 2D or 3D, the sources are distributed along the vertical direction according to Eq.~(\ref{eq:source_distribution}). As shown in the Appendix, the precise value of $\langle\Psi\rangle$ is slightly different from the approximation of Eq.~(\ref{eq:slab_Psi}). Once derived, the argument of the stable distribution~(\ref{eq:pPsi}) must be carefully rescaled to match the variable $\Psi/\langle\Psi\rangle_{sim}$ used on the horizontal axes of Fig.~\ref{fig:simu_causal_t0_a}. For completeness, the residuals between the models (2D and 3D) and the histograms are also displayed below each panel. The shaded areas correspond to the one-sigma Poissonian uncertainty coming from the histogram binning. Finally, the dashed-dotted blue line depicts the results expected from a Gaussian distribution with the same average and variance as given by the simulations. That variance is finite insofar as the light cone cut-off condition is implemented.

For energies not in excess of 10 TeV, the first noticeable result is the remarkable good convergence of the simulations toward the analytical model based on stable distributions. At low fluxes, the pdf and the left cumulative distribution of the simulations are very well matched by the theoretical 2D curve, and this holds whatever the energy considered. On the other hand, the theoretical prediction of the 3D model is always the closest to the simulations for large fluxes. Whatever the regime, all histograms reproduce the theoretical probability within {\cal O}(10\%) down to the 10$^{-4}$ level, and even with the order of magnitude below $10^{-5}$. Note that whatever the energy in the range extending from 100~GeV to 1~PeV, the simulations are not at all reproduced by the Gaussian law, featured by the dashed-dotted blue lines, which would be the limiting case for an infinite number $N$ of sources according to the conventional central limit theorem. Stable laws are, on the contrary, an excellent approximation to our results, even though a cut has been imposed on the single source pdf $p(\psi)$ from causality considerations and one would naively expect $P(\Psi)$ to relax toward a Gaussian law.

At fixed CR energy, we observe a transition occurring at some critical value $\psi_h$ of the flux $\Psi$, above which the 3D (i.e., $\alpha=5/3$) stable law yields a better approximation than the 2D (i.e., $\alpha=4/3$) distribution. In order to derive an estimate for $\psi_h$, we should keep in mind that stable laws tend to be dominated by the contribution from a single object. The transition between the 2D and 3D regimes of the total flux pdf $P(\Psi)$ should result from an evolution in the behavior of the individual flux pdf $p(\psi)$ with respect to $\psi$. This change is, in turn, related to a modification in the spatial distribution of the sources. We remark that the closer the object, the higher the flux it yields. Let us now make an educated guess and define $\psi_h$ as the critical flux above which the dominant sources are statistically very close to the observer, at a distance less than the half-thickness $h$ of the Galactic disk. As seen by the observer, they are isotropically distributed and the 3D model applies. As objects yielding a flux less than $\psi_h$ are farther, their spatial distribution reflects the flatness of the Galactic disk and the 2D model is best suited to describe the simulations. In the phase space diagram of Fig.~\ref{fig:max_psi}, the flux $\psi_h$ corresponds to the solid gray iso-flux curve where $d_c$ is replaced by $h$. A value can be derived from Eq.~(\ref{eq:psi_M}) and translates to the red solid line of Fig.~\ref{fig:Transition_2d_3d_psic} that depicts the behavior of the ratio $\psi_{h}/\langle\Psi\rangle$ as a function of CR kinetic energy. That curve
features the same trend as our results. As the CR energy increases, the 2D to 3D transition occurs at higher values of the flux relative to the average $\langle\Psi\rangle$. At low CR energy, the agreement is not very good though. The transition flux $\psi_h$ falls beneath the average flux for energies below 300~GeV, a trend that is not observed in our simulations. In the upper-left panel of Fig.~\ref{fig:simu_causal_t0_a}, the left cumulative histogram is very well matched by the 2D stable law. Furthermore, defining $\psi_h$ as the maximal flux yielded by sources located at a distance $d_c$ exactly equal to $h$ is somewhat arbitrary.

If we are interested in quantifying the probability of measuring a particular flux excess with respect to the mean, as is the case for the applications discussed in Sec.~\ref{applications}, we can alternatively define $\psi_h$ as the value for which the 2D and 3D right cumulative distributions $C(\psi)$ are equal. Following the notations of Eq.~(\ref{eq:def_cumulative}), the value of $\psi_h$ is now given by the condition ${\cal C}_{4/3}(\psi_h) = {\cal C}_{5/3}(\psi_h)$. This leads to the dashed green line of Fig.~\ref{fig:Transition_2d_3d_psic} whose behavior with respect to CR energy exhibits the same trend as the red curve of the previous estimate. This time, the transition flux is always larger than the average $\langle\Psi\rangle$. The dashed green line separates the plot in two distinct regions. In the light green domain extending below the frontier, the simulations are well explained by the theoretical 2D stable law (i.e., $\alpha=4/3$). In the light red part of the diagram, the 3D stable distribution (i.e., $\alpha=5/3$) provides the best approximation. Above 1 TeV, our new value of the transition flux is smaller than the previous estimate derived from the Galactic disk half-thickness argument. Indeed, the cumulative distribution $C(\psi)$ is obtained via an integration of the pdf $p(\psi)$ from $\psi$ upward. It contains information pertaining to the high-flux behavior of the pdf, and feels the 3D regime for smaller values of the flux compared to the other approach. Below 100~GeV, the 3D and 2D stable laws should be used above and beneath the mean flux, respectively, indicated in the plot by the solid black horizontal line.

\begin{figure*}[h!]
\centering
\includegraphics[width=1.95\columnwidth]{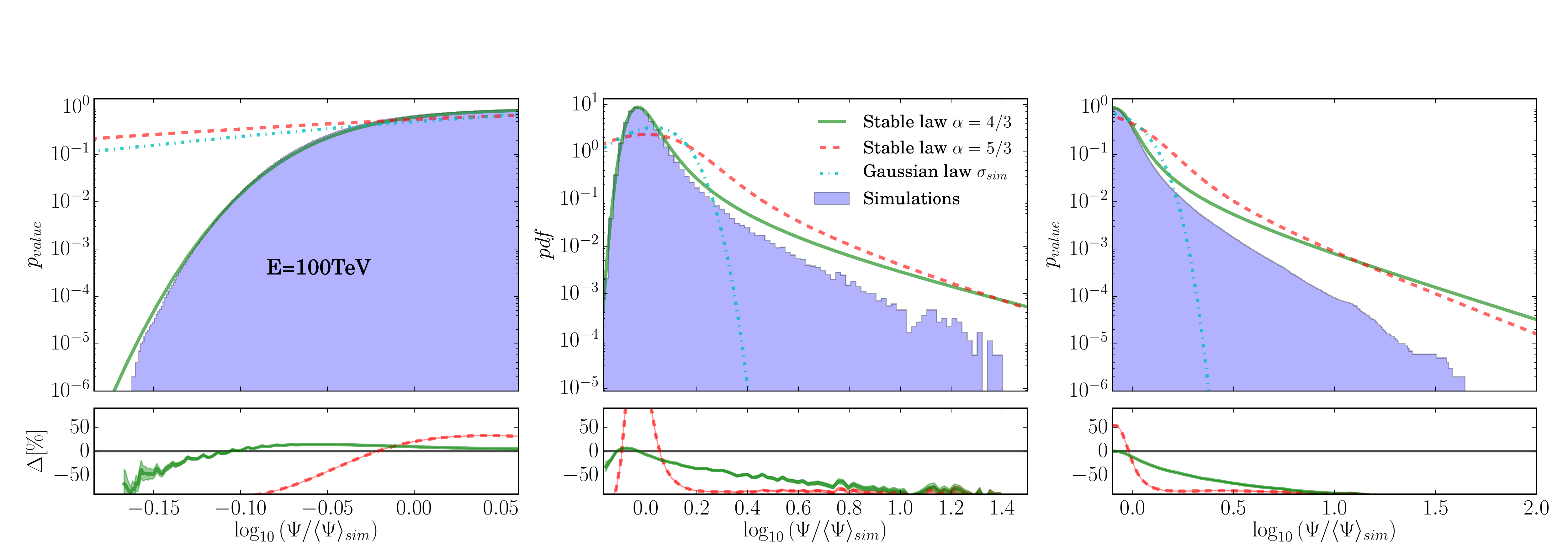}
\includegraphics[width=1.95\columnwidth]{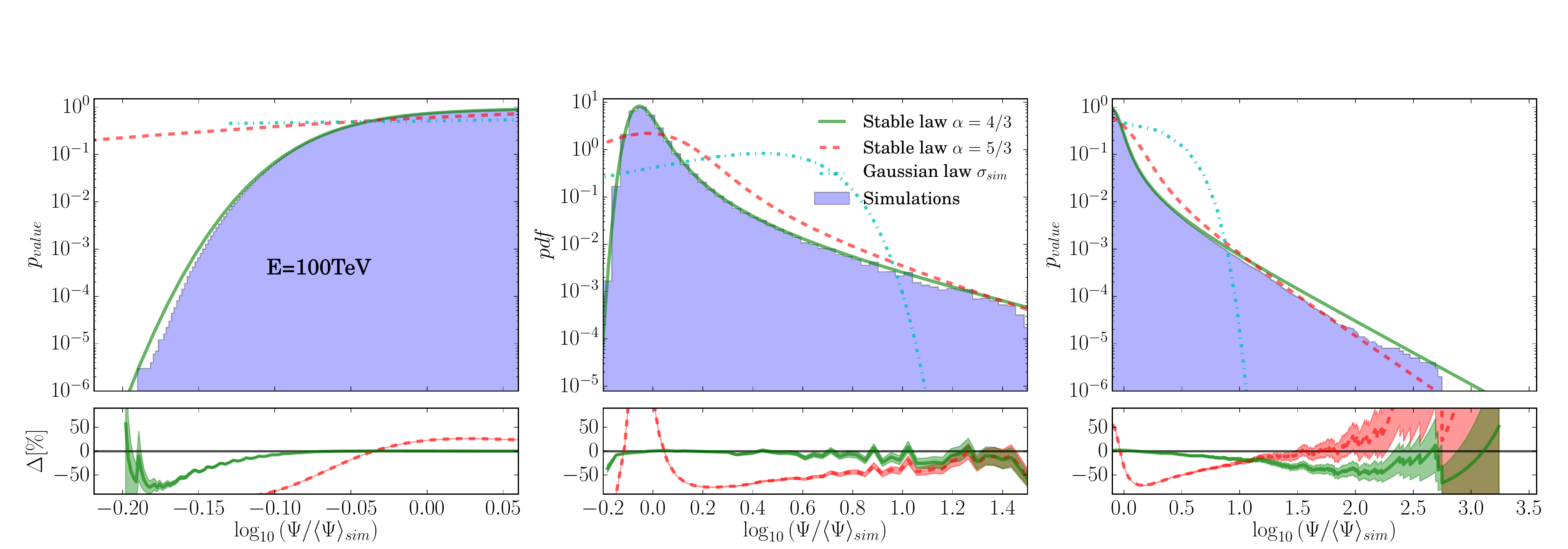}
\caption{
Same as in Fig~\ref{fig:simu_causal_t0_a} with a CR kinetic energy of 100~TeV. The upper row features the results of a simulation where the causality constraint is implemented whereas in the lower row, all sources are taken into account in the calculation of the flux $\Psi$, including those lying outside the light cone of the observer.
}
\label{fig:simu_causal_t0_b}
\end{figure*}

The shaded area in the upper-right corner of Fig.~\ref{fig:Transition_2d_3d_psic} lies above the dashed gray curve featuring the ratio $\psi_c/\langle\Psi\rangle$. In this region, causality is expected to limit the statistical excursions of the flux toward high values, and the stable law should overestimate the actual pdf $P(\Psi)$. This trend is already present in the lower-right panel of Fig.~\ref{fig:simu_causal_t0_a}, where the right cumulative blue histogram lies below the dashed red curve of the 3D stable law. For the simulations performed at 100~GeV and 1~TeV, the agreement is excellent. According to Fig.~\ref{fig:Transition_2d_3d_psic}, the light cone cut-off starts to seriously affect the (not too large) fluctuations of the flux above an energy of 10~TeV.
To illustrate this effect, we show the results of simulations realized at 100~TeV in Fig.~\ref{fig:simu_causal_t0_b}, for which the causality constraint has been switched on (upper row) or off (lower row). The behavior of the pdf $P(\Psi)$ is very different between the two cases. In the upper row, sources lying outside the light cone of the observer are removed. The discrepancy between the simulated histograms and the stable law predictions is striking, even for fluctuations only 30\% larger than the mean. If, now, all the sources are allowed to contribute to the flux $\Psi$, the agreement between the histograms and the stable laws is recovered. In the lower-right panel of Fig.~\ref{fig:simu_causal_t0_b}, the 2D prediction ${\cal C}_{4/3}$ for the right cumulative distribution accounts well for the Monte Carlo results up to a flux approximately ten times larger than the mean. For larger fluctuations, the 3D function ${\cal C}_{5/3}$ is a good match to the histograms.

One last remark is in order. According to Fig.~\ref{fig:Transition_2d_3d_psic}, the ratio $\psi_c/\langle\Psi\rangle$ drops below unity for energies above 100~TeV. In this energy range, we expect the effect of the causal cut to be so important that stable laws are poor representations of the actual pdf $P(\Psi),$ which they overshoot by a large margin. As discussed above, this occurs as soon as the flux exceeds its mean value. But for values of $\Psi$ lower than the average $\langle\Psi\rangle$, the 2D stable law prediction (solid green curve) is still in excellent agreement with the histograms, whether the light cone cut-off condition is implemented or not. In this regime, small fluxes are involved with two consequences. Distant sources dominate the faint signal received by the observer and their spatial distribution reflects the flatness of the Galactic disk, which is 2D in nature. Moreover, they yield a flux $\psi$ well below the light cone cut-off $\psi_c$ to which they are totally insensitive. Notice the excellent agreement between the simulations and the 2D stable law predictions in all the left panels of Fig.~\ref{fig:simu_causal_t0_a} and \ref{fig:simu_causal_t0_b}. We must finally conclude that for fluxes smaller than the mean, the pdf $P(\Psi)$ has asymptotically relaxed toward the (2D) stable distribution of Eq.~(\ref{eq:pPsi}), even though this is not the case in the high-flux regime. Stable laws seem to be robust descriptions of the pdf $P(\Psi)$ as soon as the condition of Eq.~(\ref{eq:SL_definition}) is fulfilled over some range of values of $\psi$.

\section{Applications}\label{applications}

In this section we present some applications of the theory developed above. At first, we want to gauge if the present precision of experimental data is sufficient to be sensitive to a fluctuation of the flux coming from the discreteness of the sources. To do so, we compute the probability that a source configuration leads to a 3$\sigma$ fluctuation above and below the average flux assumed to follow a power law spectrum in energy. We actually calculate this probability for the proton flux measured by AMS02 and PAMELA. The results are reported in Table~\ref{tab:pamVsams} for the energies of 50 GeV and 1 TeV and for the three benchmark models MIN, MED and MAX. Each box of the table corresponds to the energy given at the top of its column. The upper value is the probability of getting a fluctuation above 3$\sigma$, and the value at the bottom corresponds to the probability of having a fluctuation below 3$\sigma$.
The first noticeable feature is that the probability of seeing a 3$\sigma$ fluctuation above the mean is always different than a 3$\sigma$ fluctuation below. This is a consequence of the huge asymmetry of the stable distribution. Fluctuations below the mean are strongly prevented below the brutal fallout of the pdf. Furthermore, the probability of measuring a 3$\sigma$ deviation from the mean is always larger at 50 GeV than at 1 TeV. This result means that the experimental uncertainty increases faster than the typical spread of the stable law with the energy. Thus, by improving data precision thanks to higher statistics or a new experiment, the probability of observing deviation from the mean flux at \textit{low} energies ($\approx$50 GeV) increases. Regarding the different propagation models, we notice that a fluctuation in the MAX model with a large halo height is much less expected than a fluctuation for small halo models, to which MIN belongs. In other words, the fact that AMS02 proton data do not show any departure from a power law spectrum at relatively low energies can be interpreted as an independent hint for large halo size models. Finally, comparing PAMELA results with AMS02, we notice that the latter has made a large step forward in reducing the experimental uncertainties, giving hope to chance of seeing deviation of the power law independently from  the propagation models. We note that the effect of the stochasticity of the sources is expected to be smaller when dealing with secondary nuclei, since the interstellar gas on which they are produced is expected to be more smoothly distributed than Supernova remnants. Secondary to primary ratios are sensitive to this difference and may lead to biased results when extracting propagation parameters.

\begin{table*}
        \centering
\begin{tabular}{l ||c| c|| c| c }
\hline
\multicolumn{1}{c}{Models}& \multicolumn{2}{c}{PAMELA} & \multicolumn{2}{c}{AMS02}  \\
\hline
\multirow{3}*{Model}& 50GeV &  1TeV                 & 50GeV  & 1TeV   \\[0.5mm]
                                        & $p\left(\Psi>\langle\Psi\rangle+3 \sigma\right)$         & $p\left(\Psi>\langle\Psi\rangle+3 \sigma\right)$ & $p\left(\Psi>\langle\Psi\rangle+3 \sigma\right)$ & $p\left(\Psi>\langle\Psi\rangle+3 \sigma\right)$ \\ [2mm]
                                        & $p\left(\Psi<\langle\Psi\rangle-3 \sigma\right)$         & $p\left(\Psi<\langle\Psi\rangle-3 \sigma\right)$ & $p\left(\Psi<\langle\Psi\rangle-3 \sigma\right)$ & $p\left(\Psi<\langle\Psi\rangle-3 \sigma\right)$ \\\hline
\multirow{2}*{MIN}       & 0.15         & 0.083       & 0.28          & 0.26 \\ [2mm]
                                        & 0.13         & $<10^{-6}$  & 0.63          & 0.51\\
\hline
\multirow{2}*{MED}       & 0.047        & 0.014       & 0.16          & 0.12 \\ [2mm]
                                        & $<10^{-6}$   & $<10^{-6}$  & 0.26          & 0.0025\\
\hline
\multirow{2}*{MAX}       & 0.009        & 0.0018      & 0.045         & 0.016\\ [2mm]
                                        & $<10^{-6}$   & $<10^{-6}$  & $<10^{-6}$    & $<10^{-6}$\\
\hline
\end{tabular}

\vspace{0.7cm}

\caption{Probability that a source configuration leads to a 3$\sigma$ fluctuation above and below the flux measured by AMS02 and PAMELA. The calculation is made for the three benchmark propagation models MIN, MED, and MAX, and for the two energies 50 GeV and 1 TeV.}\label{tab:pamVsams}
\end{table*}

A positive large flux fluctuation corresponds to the situation where some of the sources are very near and very young. The extreme case for which a few sources (or even one source) dominate the contribution to the flux has been considered, for instance, in~\citetads{2015PhRvL.115r1103K}: the authors suggest explaining the low energy proton flux below $\sim {\cal O}(10)\,$TeV by involving the major contribution of a local SNR (within a few hundreds of pc), which exploded approximately 2 Myr ago. According to Fig. 1 of this paper, this contribution would overcome the mean flux by a factor 2.86 at the energy of $10^{3}$\;GeV. In our myriad model, which assumes isotropic diffusion, the probability that a peculiar configuration of sources leads to a deviation comparable or larger to the one stated in this paper, is given by:\\
\begin{align}
p_s & =\int_{2.86\langle\Psi\rangle}^{\infty}P(\Psi)\;\mathrm{d}\Psi \nonumber \\ & =1-\int_{0}^{2.86\langle\Psi\rangle}P(\Psi)\;\mathrm{d}\Psi\nonumber \\ &=1-\underbrace{(1+\epsilon)^{N}}_{\approx1\text{ (here)}}\int^{\frac{2.86\langle\Psi\rangle}{\sigma_N}}_{0}\;S[\alpha,1,1,0;1](X)\;\mathrm{d}X\;.\label{eq:proba_application}
\end{align}
Such a deviation corresponds to $\log_{10}(\psi/\langle\Psi\rangle)\approx0.46$ at $10^{3}$\;GeV, for which Fig.~\ref{fig:Transition_2d_3d_psic} recommends the use of the 3D case corresponding to $\alpha=5/3$. This conclusion holds for the MED model for which Fig.~\ref{fig:Transition_2d_3d_psic} was made, however we checked that it was actually also the case for the MIN and MAX cases. For the MIN case, $\psi/\langle\Psi\rangle$ also falls below the condition $\psi_{max}/\langle\Psi\rangle$ as shown in Fig.~\ref{fig:ratio_psi_c_psi_env}.

The probabilities for the three different benchmark propagation models are reported in Tab.~\ref{tab:prob_table}. For comparison, we also display the probabilities obtained by using a Gaussian law with the variance of the simulations. In the homogeneous diffusion framework, this result suggests that the chance probability for such an excursion is at most at the level of $\sim 0.1\%$, and even one order of magnitude smaller if the MAX model, apparently closer to the recent observations, is adopted. It would not be correct to discard the model in~\citetads{2015PhRvL.115r1103K} based on these considerations, however, since in that article the authors advocate a strongly anisotropic diffusion. Certainly, it emphasizes the importance of this ingredient in the plausibility of the scenario.
\begin{table}
        \centering
\begin{tabular}{l c c c}
\hline
Models & MIN & MED& MAX \\
\hline \hline
        Probabilities (Stable law 5/3) & 0.0072 & 0.0012 & 0.00016\\ [2mm]
\hline
Probabilities (Gaussian law) & 0.06 & $10^{-5}$ & 0\\ [2mm]
        \hline
\end{tabular}
\caption{Probabilities of obtaining a flux larger than $2.86\langle\Psi\rangle$ at 1 TeV in the myriad model, calculated for three benchmark propagation models MIN, MED, and MAX. The Gaussian probability is extracted from the simulation and crucially depends on the integration time of the simulation.}\label{tab:prob_table}
\end{table}

Another example is provided by the scenario discussed in \citetads{2015ApJ...803L..15T}. Here the authors invoke a two components model for which the high energy CR spectrum is dominated by the average Galactic population, and the low energy part by one local old source, or, alternatively, a population
of local old sources. In this case, homogeneous diffusion is assumed. The two different energy dependences of these components would explain the break in the proton and helium flux above 200-300 GV. Once more, we can compute the probability for such a low-energy fluctuation of the flux in our myriad model, assuming the mean flux to be reached above the spectral break. From  Fig. 2 of this paper, the proton flux at $E=10$ GeV is dominated by some local sources, which yield a value of $\Psi$ approximately 3.3 times the average $\langle\Psi\rangle$. Within their propagation model, one can show that the probability of such an excess must be treated with the 3D case. Making use of the formula in Eq.~(\ref{eq:proba_application}) of the previous example, we obtain a probability of $8.6\times 10^{-5}$.
Thus we can conclude that the only reasonable possibility for their scenario to be true is to assume a sum of two populations of sources, with the observed flux at the Earth being close to the sum of their  average contributions rather than due to a local fluctuation.

\begin{figure*}[h!]
\centering
\includegraphics[width=1.07\columnwidth]{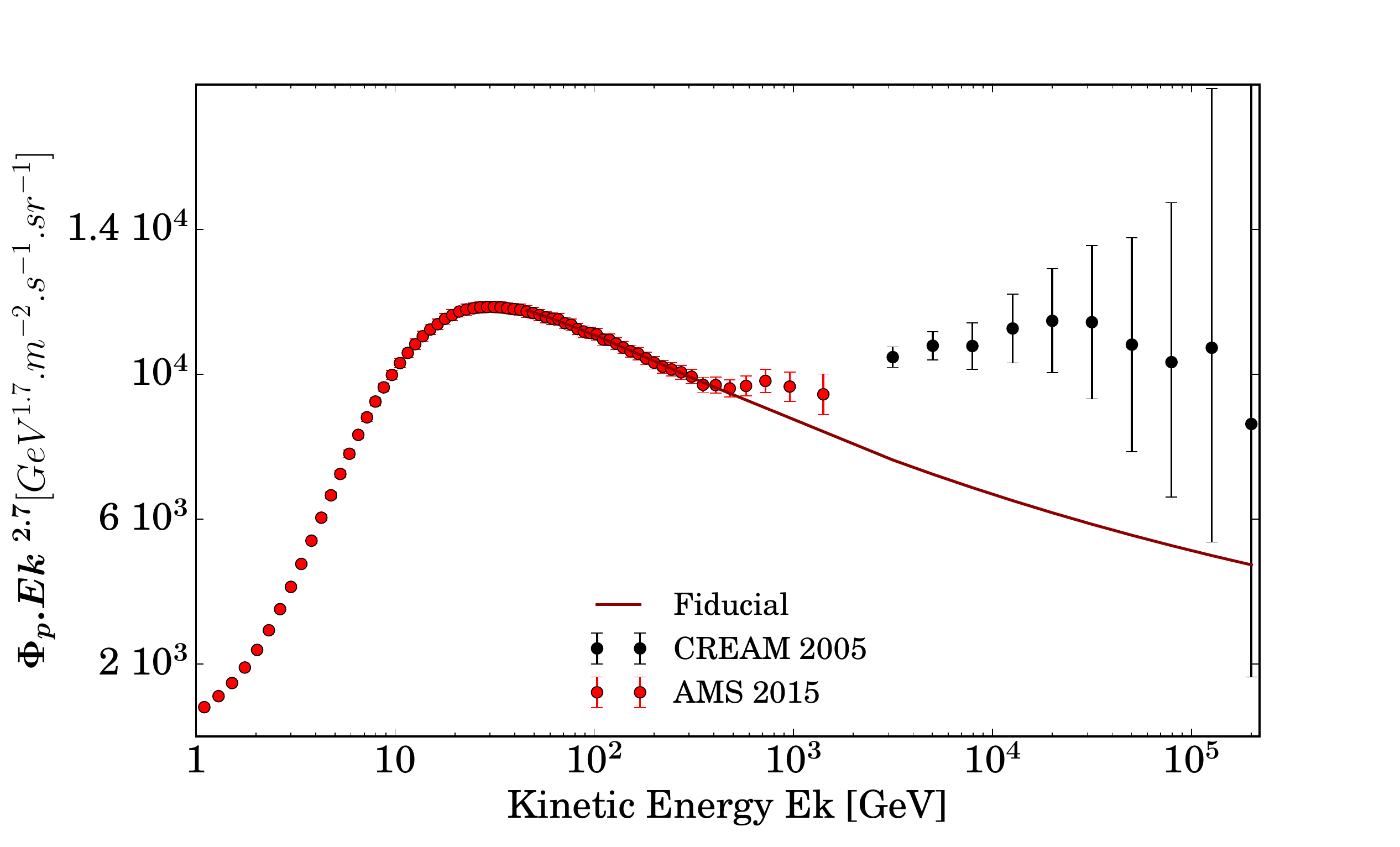}
\includegraphics[width=0.93\columnwidth]{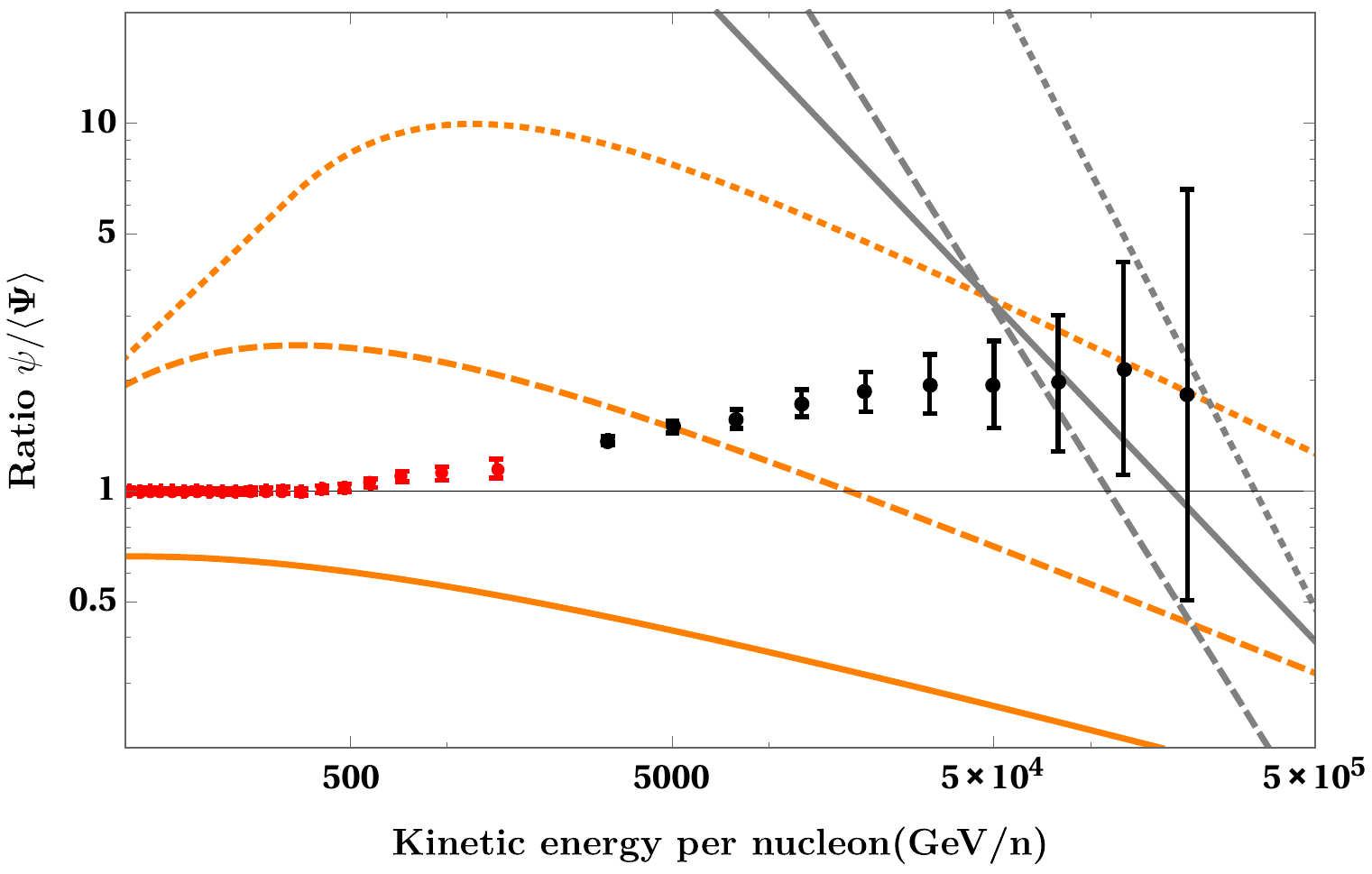}
\caption{Left panel: proton flux from AMS02 \citepads{2015PhRvL.114q1103A} and CREAM \citepads{2011ApJ...728..122Y}, and a fit of the spectrum between 45 GeV and 200 GeV that we assume here to be the mean Galactic flux. Right panel: data divided by the theoretical mean above 45 GeV, together with conditions $\psi_c/\langle\Psi\rangle$ and $\psi_{max}/\langle\Psi\rangle$ of Fig.~\ref{fig:ratio_psi_c_psi_env} (solid for MAX model, dashed for MED, and dotted for MIN).}
\label{fig:Proton_Mean}
\end{figure*}

Finally, one may consider the opposite possibility (advanced, for instance, in~\citeads{2012A&A...544A..92B,2013A&A...555A..48B}) for which the high-energy flux is a signature of the contribution of local sources, while the steeper flux at lower energies follows the Galactic average. In the left-hand panel of Fig.~\ref{fig:Proton_Mean} we display the inferred mean proton flux in the range [45-200] GeV in this model, from which data depart more and more above the energies $\gtrsim$200 GeV. To estimate the probability that such a discrepancy may occur, it is crucial to check the requirement for the applicability of the stable law, that is, $\Psi<\psi_c$. In the right-hand panel of Fig.~\ref{fig:Proton_Mean}, we plot the data divided by the mean above 45 GV, together with conditions $\psi_c/\langle\Psi\rangle$ and $\psi_{max}/\langle\Psi\rangle$ of Fig.~\ref{fig:ratio_psi_c_psi_env} (solid for MAX model, dashed for MED, and dotted for MIN). If the data fall above the gray lines, it means that the observed excess cannot be provided by local sources {in the diffusive regime}. This is what happens to the two (three) highest energy CREAM data in the MED (MAX) propagation model. Strictly speaking, we can only conclude that our theory is inapplicable to those energies in the framework of these propagation models, since the diffusion approximation breaks down. However, it also means that the only way one or a few local sources might account for the measured flux in that range is to assume that CR propagate quasi-ballistically from the hypothetical source(s), which would qualitatively lead to ${\cal O}(1)$ anisotropy, in blatant contrast with the data, showing a dipole anisotropy in this energy range at or below the 0.1\% level (see, e.g., the compilation in Fig. 2 of~\cite{Blasi:2011fm}). We also note that the data fall above the orange lines at all energies for the MAX model and already at 5 TeV for the MED model. These lines correspond to the ``maximal excursion'' due to a source as young as $\tau<\tau_c=2.7\rm \,kyr$ and as close as $r<d_c=0.06\,\rm kpc$, which is approximately the closest in space-time estimated on the basis of the available catalog. Although the available catalog may be incomplete, it is less and less likely the case for close and young/powerful sources. This is another independent argument suggesting that a local explanation for the high-energy break of the type invoked in~\citeads{2012A&A...544A..92B,2013A&A...555A..48B} is unlikely in propagation scenarios of the MED or MAX type. If we discard all constraints from catalogs, our theory is applicable below 50 TeV to estimate the probability of such an excess within our myriad model. The probability is calculated as
\beq
p_{value}=\int_{\Psi_{exp}}^{\infty}d\psi_{exp}\int_{0}^{+\infty}\;d\psi\;p(\psi_{exp}|\psi)\;p(\psi|myriad)\;,
\eeq
where $p(\psi_{exp}|\psi)$ is a Gaussian law of spread, $\sigma_{exp}$ the experimental variance, and $p(\psi|myriad)$ is the probability of achieving a theoretical flux $\psi$ in the myriad model. We compute this probability for the most constraining data point, which lies at 12.6 TeV in the CREAM data. The fluctuation at this energy is $\psi/\langle\Psi\rangle\approx1.73,$ which justifies the use of the 2D case with the stable law $\alpha=4/3$. The results are reported in Table.~\ref{tab:prob_table2}. We obtain a maximum of 3\%\ within the MIN scenario. This probability is small but not vanishingly small. In fact, independent CR arguments disfavoring the MIN scenario (see for instance~\cite{Lavalle:2014kca,Giesen:2015ufa,Kappl:2015bqa,Evoli:2015vaa}) are probably even more capable of providing a killing blow to this model.

\begin{table}
        \centering
\begin{tabular}{l c c c}
\hline
Models & MIN & MED& MAX \\
\hline \hline
        Probabilities(Stable law 4/3) & 0.031 & 0.0082 & 0.0013\\ [2mm]
        \hline
\end{tabular}
\caption{Probabilities calculated for the most discriminating point of CREAM at 12.6TeV (the fourth one) and the three benchmark propagation models.}\label{tab:prob_table2}
\end{table}

\section{Conclusions}\label{conclusions}
 Given the precision currently reached by cosmic ray measurements, it is more and more important to assess uncertainties associated
 with different theoretical predictions. The space-time discreteness of the cosmic ray sources is an important cause
 of theoretical uncertainty, given the the lack of information available on their precise epochs and locations, with the possible exception of the most recent and close ones.

In this article we have elaborated a statistical theory to deal with this problem, relating the composite probability $P(\Psi)$ to obtain a flux $\Psi$ at the Earth
 and to the single-source probability $p(\psi)$ to contribute to a flux $\psi$. The main difficulty arises since $p(\psi)$ is a ``heavy tail'' distribution, characterized by power-law or broken power-law behavior up to very large fluxes for which the central limit theorem does not hold, and leading to peculiar function, stable under the convolution; namely stable laws different from the Gaussian distributions.

We have analytically discussed the regime of validity of the stable laws associated with the distributions arising in cosmic ray astrophysics for different propagation parameters and energy ranges, as well as the limitations to the treatment imposed by causal considerations and partial source catalog knowledge. We have also validated our results with extensive Monte Carlo simulations.

We find that relatively simple recipes provide a remarkably satisfactory description of the probability $P(\Psi)$. We also find that a naive Gaussian fit to simulation
results would underestimate the probability of very large fluxes  several times above the average, while overestimating the probability of
relatively milder excursions. At large energies, large flux fluctuations are prevented by causal considerations, while at low energies, a partial knowledge
of the recent and nearby population of sources plays an important role.

We have applied our theory to some models recently discussed in the literature attempting to explain the spectral breaks as effects of a prominent nearby
source. We showed that, at least within homogeneous and isotropic diffusion models, it is unlikely that this is the cause of the observed phenomenon, since
the only case where this might happen with an appreciable probability is disfavored by independent arguments involving  secondary tracers such as positrons,
antiprotons, and/or  the boron/carbon CR spectrum.
We have also argued that the precision recently attained by cosmic ray measurements makes the observation of upward departures from the mean
expectations more likely. Actually, the close agreement of recent AMS-02 at relatively low-energies with average expectations of continuous cosmic ray source models
represents by itself a constraint on propagation models, which intriguingly goes in the same direction as those recently derived from secondary species. Diffusion models
with a large halo and mild energy dependence appear favored.  Another theoretically robust prediction is that no significant downward fluctuation
with respect to average model expectations should be observed, a fact that, for the time being, seems to be confirmed by the data.

The formalism elaborated and validated in this article constitutes only a first step of a potentially much broader program. A trivial extension of the
theory allows one to deal with several uncorrelated populations of sources, each one with its own distribution. One may also apply this
formalism in a slightly modified form to deal with effects on secondary nuclei produced onto inhomogeneous medium with ``heavy-tail'' inhomogeneity distribution probability.
A more subtle generalization would be required to deal with correlations of flux predictions at different energies. Even more challenging is to elaborate an analytical theory accounting for  space and time correlations  among the discrete sources of cosmic rays. Last but not least, it might be worth entertaining the possibility that some of the tools developed for applications
to cosmic ray flux problems may find an application in other contexts of astroparticle physics, if not of physics in general.

%
\vskip 1.0cm
\begin{acknowledgements}
We are grateful to our colleague P. Briand from the laboratory of mathematics LAMA of Universit\'e de Savoie Mont-Blanc for his kind
help and for having provided us with a proof of the generalized central limit theorem. We also thank P. Mertsch who contributed by
drawing our attention to the role of stable laws in cosmic ray physics.
\end{acknowledgements}
\appendix


\section{The MIN, MED, and MAX models}\label{annexe:0}

In Table~\ref{tab:model:annexe} we recall the values of benchmark propagation models used in \citeads{2004PhRvD..69f3501D}. Note that for our concerns, only $K_0$, $\delta$ and $L$ are relevant, since by virtue of the generalised central limit theorem, convection and reacceleration do not affect the shape of the distribution but only its mean.
\begin{table*}
\centering
\begin{tabular}{c c c c c c }
\hline
Model & $\delta$ & $K_0{\rm [kpc^2/Myr]}$ & $L$[kpc] & $V_c$[km/s] & $V_a$[km/s]\\
\hline
\hline
MIN & 0.85 & 0.0016 & 1  & 13.5 & 22.4\\
MED & 0.70 & 0.0112 & 4  & 12   & 52.9\\
MAX & 0.46 & 0.0765 & 15 & 5    & 117.6\\

\hline
\end{tabular}

\vspace{0.7cm}

\caption{Propagation parameters from the MIN, MED, and MAX models as chosen in \citeads{2004PhRvD..69f3501D}}\label{tab:model:annexe}
\end{table*}

\section{What do we mean by mean}\label{annexe:1}

The mean flux of cosmic rays is a very useful observable whose theoretical derivation is relatively simple in the high-energy limit where diffusion is the dominant propagation mechanism. Two simplifications are commonly used in the literature and lead to very good approximations compared to more sophisticated models.

To commence, sources are assumed to lie within an infinite plane with half-thickness $h$ sandwiched by two larger diffusion volumes with height $L$. Inside this magnetic halo, propagation is characterized by the diffusion coefficient $K$. The so-called \textit{infinite slab} model requires in addition to consider the disk as infinitesimally thick. In the steady state regime, the total mean flux satisfies the equation
\beq
-K\,\nabla^2\langle\Psi\rangle_{slab}= 2\,h\,\delta(z)\;Q \;.
\eeq
Assuming that the CR density vanishes at the vertical boundaries $z = \pm L$, one readily gets
\beq
\langle\Psi\rangle_{slab}=Q\;\frac{h\,L}{K}\equiv Q\;\tau_D \quad\text{     with   }\quad [Q]=[{\rm time}]^{-1} [\Psi] \;.
\label{eq:A_slab}
\eeq
In this equation, $\Psi$ is homogeneous to the density of cosmic rays expressed in particles per unit of energy and of volume or, equivalently for the discussion, to the flux expressed in particles per unit of energy, time, surface, and solid angle. A simple rescaling by the factor $v_{CR}/4\pi$, where $v_{CR}$ is the cosmic ray velocity, can be applied to switch from one quantity to the other. Choosing the former, the injection rate $Q$ is interpreted as the number of particles injected per unit of energy, time and volume in the Galaxy. It may be useful to factorize $Q= q\,\nu / V_{\rm MW}$. Here, the spectrum $q$ of the particles injected by a single source is expressed in particles per unit of energy. Sources appear with a rate $\nu$. Assuming these are supernova remnants implies a value of three SN explosions per century. Finally, the Galactic disk with half-thickness $h$ and radius $R$ encompasses a volume $V_{\rm MW}=2\,\pi\,h\,R^2$.

To go a step forward, we may now assume that the sources are no longer pinched inside an infinitesimally thick Galactic disk, but are spread over a vertical distance of $2h$. As long as steady state holds, the total mean flux now satisfies the equation
\beq
-K\,\nabla^2\langle\Psi\rangle_{vol}= \;\Theta (h-|z|)\; Q \;,
\eeq
whose solution, derived with the same vertical boundary conditions as previously, may be expressed as
\beq
\langle\Psi\rangle_{vol}=\langle\Psi\rangle_{slab}\;\left(1-\frac{h}{2\,L}\right)\;.
\label{eq:A_vol}
\eeq
Sources extend now along the vertical direction and are no longer packed at $z=0$. They yield a flux $\langle\Psi\rangle_{vol}$ slightly smaller than $\langle\Psi\rangle_{slab}$.

The theoretical expression of the probability $P(\Psi)$ is provided by the stable law of Eq.~(\ref{eq:pPsi}). Its argument depends on the average value $\langle\Psi\rangle_{th}$ which should be consistently derived within the Milky Way model to which the pdf $P(\Psi)$ is associated. Although expressions~(\ref{eq:A_slab}) and (\ref{eq:A_vol}) are excellent approximations to the theoretical mean $\langle\Psi\rangle_{th}$, they should not be used.
In particular, the solid green (2D) and dashed red (3D) curves of Fig.~\ref{fig:simu_causal_t0_a} and \ref{fig:simu_causal_t0_b} are based on the assumption that the Galactic disk has radius $R$ and that sources cannot be older than $T = 3 \tau_0$. Furthermore, all sources contribute to the theoretical average $\langle\Psi\rangle_{th}$, including those lying outside the light cone of the observer. Causality is not implemented and the heavy tail behavior, which the theoretical pdf $p(\psi)$ should exhibit, is not suppressed.

In order to compare the theoretical pdf $P(\Psi)$ with the simulations, $\langle\Psi\rangle_{th}$ needs to be calculated from the convolution of the source term ${\cal Q}({\bf x}_S,t_S)$ with the diffusive propagator ${\cal G}_B$ over the volume of space-time ${\cal V}$ covered by the simulation
\beq
{\cal V}=V_{\rm MW} \times 3 \tau_0 \quad\text{     where   }\quad \tau_0 \equiv \frac{\pi}{4} \, {\displaystyle \frac{L^{2}}{K}} \;.
\eeq
The theoretical average corresponding to the 2D case, where the sources are pinched inside an infinitesimally thick disk, is denoted by $\langle\Psi\rangle_{2D,~R,~T=3\tau_0}$ whereas the result corresponding to the 3D case, for which the sources are vertically spread over a distance $2h$, is denoted $\langle\Psi\rangle_{3D,~R,~T=3\tau_0}$. These quantities are calculated as follows

\begin{align}
\left. \begin{array}{lr} \langle\Psi\rangle_{3D,~R,~T=3\tau_0} \\ \\ \langle\Psi\rangle_{2D,~R,~T=3\tau_0} \end{array} \right\} =
\frac{q\,\nu }{V_{\rm MW}} &\times {\displaystyle \int_{0}^{3\tau_0}} \! d\tau_S \;
{\displaystyle \int_{0}^{R}} \! 2 \, \pi \, r_{S} dr_{S} \,
\left\{{\displaystyle \frac{\exp\left({-r_{S}^{2}}/{4 K \tau_{S}}\right)}{4 \, \pi \, K \, \tau_{S}}}\right\} \;\nonumber\\
&\times {\displaystyle \int_{-h}^{h}} dz_{S} \, {\cal V}_{B}(z_{S},\tau_S) \times
\begin{cases}1&\text{in 3D, or}\\&\\2h\,\delta(z_{S})&\text{in 2D.}\end{cases}
\end{align}
The function ${\cal V}_{B}(z_{S},\tau_S)$ describes the CR vertical propagation and takes into account the boundary conditions at $z=\pm L$. It gauges the contribution at the observer located at $z=0$ from a source that exploded a time $\tau_{S}$ ago at $z=z_{S}$. It can be expressed as the series
\begin{align}
&{\cal V}_{B}(z_{S},\tau_S) = {\displaystyle \sum_{n = - \infty}^{+ \infty}} \; (-1)^{n} \;
{\displaystyle \frac{\exp\left({-z_{n}^{2}}/{4 K \tau_{S}}\right)}{\sqrt{4 \, \pi \, K \, \tau_{S}}}}
\;\; \\&\text{where the nth image is located at} \;\;
z_{n} = 2 \, L \, n \, + (-1)^{n} \, z_{S} \;.\nonumber
\end{align}
We notice that to recover the average of the slab model, the integral for the 2D case needs to be extended to an infinite age $T$ and Galactic radius $R$. Hence, we may write $\langle\Psi\rangle_{slab}\equiv\langle\Psi\rangle_{2D,~R=\infty,~T=\infty}$. When the same integration limits are taken for the 3D case, we recover the expression of $\langle\Psi\rangle_{vol}$, which may also be defined as $\langle\Psi\rangle_{3D,~R=\infty,~T=\infty}$.

In Table~\ref{tab:mean:annexe}, we report the average fluxes calculated with the aproaches discussed above, and compare them with the simulations. Below 10~TeV, all the values are very close to each other within ${\cal O}(1\%)$. Above that energy, the mean from the simulations $\langle\Psi\rangle_{sim}$ becomes significantly lower than the theoretical one when the light cone cut-off is imposed on simulations. This effect can be checked by calculating the average flux $\langle\Psi\rangle_{sim}^{NC}$ yielded by numerical simulations for which no causality constraint has been imposed. We get results that are always within the one sigma Poissonian error of theoretical one $\langle\Psi\rangle_{3D,~R,~T=3\tau_0}$.

\begin{table*}
\centering
\begin{tabular}{c c c c c c c }
\hline
Ek (GeV)&  $\langle\Psi\rangle_{slab}\equiv\langle\Psi\rangle_{2D,~R=\infty,~T=\infty}$ & $\langle\Psi\rangle_{2D,~R,~T=3\tau_0}$ & $\langle\Psi\rangle_{vol}\equiv\langle\Psi\rangle_{3D,~R=\infty,~T=\infty}$ & $\langle\Psi\rangle_{3D,~R,~T=3\tau_0}$ & $\langle\Psi\rangle_{sim}$ & $\langle\Psi\rangle_{sim}^{NC}$ \\
\hline
\hline
$10^{2}$ & 1.605e-01  & 1.599e-01 & 1.585e-01 & 1.579e-01 & 1.580e-01 & 1.579e-01 \\
$10^{3}$ & 2.070e-04  & 2.062e-04 & 2.044e-04 & 2.037e-04 & 2.036e-04 & 2.037e-04 \\
$10^{4}$ & 2.612e-07  & 2.601e-07 & 2.579e-07 & 2.571e-07 & 2.554e-07 & 2.577e-07 \\
$10^{5}$ & 3.289e-10  & 3.270e-10 & 3.248e-10 & 3.236e-10 & 3.075e-10 & 3.241e-10 \\
$10^{6}$ & 4.141e-13  & 4.105e-13 & 4.089e-13 & 4.072e-13 & 2.923e-13 & 4.071e-13 \\
\hline
\end{tabular}

\vspace{0.7cm}

\caption{In this table we report the theoretical average of the flux calculated within the slab model $\langle\Psi\rangle_{slab}$, the slab model taking into account the thickness of the source disk $\langle\Psi\rangle_{vol}$, the 2D model in the conditions of the simulations $\langle\Psi\rangle_{2D,~R,~T=3\tau_0}$, and the 3D model in the conditions of the simulations $\langle\Psi\rangle_{3D,~R,~T=3\tau_0}$. We also show the average of the flux obtained from the simulations without the contribution of non-causal sources $\langle\Psi\rangle_{sim}$, and with their contributions $\langle\Psi\rangle_{sim}^{NC}$. The values are given in units of $\rm[q_0.kpc^{-2}.str^{-1}]$ for the kinetic energies probed by the simulations.  }\label{tab:mean:annexe}
\end{table*}

\bibliographystyle{aa}
\bibliography{draft}

\begin{thebibliography}{24}
\expandafter\ifx\csname natexlab\endcsname\relax\def\natexlab#1{#1}\fi

\bibitem[{{Aguilar} {et~al.}(2015){Aguilar}, {Aisa}, {Alpat}, {Alvino},
  {Ambrosi}, {Andeen}, {Arruda}, {Attig}, {Azzarello}, {Bachlechner}, \&
  et~al.}]{2015PhRvL.114q1103A}
{Aguilar}, M., {Aisa}, D., {Alpat}, B., {et~al.} 2015, Physical Review Letters,
  114, 171103

\bibitem[{{Ahlers} {et~al.}(2013){Ahlers}, {Anchordoqui}, \&
  {Taylor}}]{2013PhRvD..87b3004A}
{Ahlers}, M., {Anchordoqui}, L.~A., \& {Taylor}, A.~M. 2013, \prd, 87, 023004

\bibitem[{{Bernard} {et~al.}(2013){Bernard}, {Delahaye}, {Keum}, {Liu},
  {Salati}, \& {Taillet}}]{2013A&A...555A..48B}
{Bernard}, G., {Delahaye}, T., {Keum}, Y.-Y., {et~al.} 2013, \aap, 555, A48

\bibitem[{{Bernard} {et~al.}(2012){Bernard}, {Delahaye}, {Salati}, \&
  {Taillet}}]{2012A&A...544A..92B}
{Bernard}, G., {Delahaye}, T., {Salati}, P., \& {Taillet}, R. 2012, \aap, 544,
  A92

\bibitem[{Blasi \& Amato(2012)}]{Blasi:2011fm}
Blasi, P. \& Amato, E. 2012, JCAP, 1201, 011

\bibitem[{{Delahaye} {et~al.}(2010){Delahaye}, {Lavalle}, {Lineros}, {Donato},
  \& {Fornengo}}]{2010A&A...524A..51D}
{Delahaye}, T., {Lavalle}, J., {Lineros}, R., {Donato}, F., \& {Fornengo}, N.
  2010, \aap, 524, A51

\bibitem[{{Donato} {et~al.}(2004){Donato}, {Fornengo}, {Maurin}, {Salati}, \&
  {Taillet}}]{2004PhRvD..69f3501D}
{Donato}, F., {Fornengo}, N., {Maurin}, D., {Salati}, P., \& {Taillet}, R.
  2004, \prd, 69, 063501

\bibitem[{Evoli {et~al.}(2015)Evoli, Gaggero, \& Grasso}]{Evoli:2015vaa}
Evoli, C., Gaggero, D., \& Grasso, D. 2015, JCAP, 1512, 039

\bibitem[{Giesen {et~al.}(2015)Giesen, Boudaud, G{\'e}nolini, Poulin, Cirelli,
  Salati, \& Serpico}]{Giesen:2015ufa}
Giesen, G., Boudaud, M., G{\'e}nolini, Y., {et~al.} 2015, JCAP, 1509, 023

\bibitem[{{Higdon} \& {Lingenfelter}(2003)}]{2003ApJ...582..330H}
{Higdon}, J.~C. \& {Lingenfelter}, R.~E. 2003, \apj, 582, 330

\bibitem[{{Kachelrie{\ss}} {et~al.}(2015){Kachelrie{\ss}}, {Neronov}, \&
  {Semikoz}}]{2015PhRvL.115r1103K}
{Kachelrie{\ss}}, M., {Neronov}, A., \& {Semikoz}, D.~V. 2015, Physical Review
  Letters, 115, 181103

\bibitem[{Kappl {et~al.}(2015)Kappl, Reinert, \& Winkler}]{Kappl:2015bqa}
Kappl, R., Reinert, A., \& Winkler, M.~W. 2015, JCAP, 1510, 034

\bibitem[{Lagutin \& Nikulin(1995)}]{lagutin1995fluctuations}
Lagutin, A. \& Nikulin, Y.~A. 1995, Soviet Journal of Experimental and
  Theoretical Physics, 81, 825

\bibitem[{Lavalle {et~al.}(2014)Lavalle, Maurin, \& Putze}]{Lavalle:2014kca}
Lavalle, J., Maurin, D., \& Putze, A. 2014, Phys. Rev., D90, 081301

\bibitem[{Lee(1979)}]{lee1979statistical}
Lee, M. 1979, The Astrophysical Journal, 229, 424

\bibitem[{L{\'e}vy(1925)}]{levy1925calcul}
L{\'e}vy, P. 1925, Calcul des probabilit{\'e}s (Gauthier-Villars)

\bibitem[{Mandelbrot(1960)}]{mandelbrot1960pareto}
Mandelbrot, B. 1960, International Economic Review, 1, 79

\bibitem[{Mertsch(2011)}]{Mertsch:2010fn}
Mertsch, P. 2011, JCAP, 1102, 031

\bibitem[{Nolan(2012)}]{nolan2012stable}
Nolan, J.~P. 2012, Stable distributions, Vol. 1177108605 (ISBN)

\bibitem[{Serpico(2015)}]{Serpico:2015caa}
Serpico, P.~D. 2015, in {Proceedings, 34th International Cosmic Ray Conference
  (ICRC 2015)}

\bibitem[{{Tomassetti} \& {Donato}(2015)}]{2015ApJ...803L..15T}
{Tomassetti}, N. \& {Donato}, F. 2015, \apjl, 803, L15

\bibitem[{Uchaikin \& Zolotarev(1999)}]{uchaikin1999chance}
Uchaikin, V.~V. \& Zolotarev, V.~M. 1999, Chance and stability: stable
  distributions and their applications (Walter de Gruyter)

\bibitem[{{Wolfram Research, Inc.}(2016)}]{mathematica}
{Wolfram Research, Inc.} 2016, Mathematica 11.0

\bibitem[{{Yoon} {et~al.}(2011){Yoon}, {Ahn}, {Allison}, {Bagliesi}, {Beatty},
  {Bigongiari}, {Boyle}, {Childers}, {Conklin}, {Coutu}, {DuVernois}, {Ganel},
  {Han}, {Jeon}, {Kim}, {Lee}, {Lutz}, {Maestro}, {Malinine}, {Marrocchesi},
  {Minnick}, {Mognet}, {Nam}, {Nutter}, {Park}, {Park}, {Seo}, {Sina},
  {Swordy}, {Wakely}, {Wu}, {Yang}, {Zei}, \& {Zinn}}]{2011ApJ...728..122Y}
{Yoon}, Y.~S., {Ahn}, H.~S., {Allison}, P.~S., {et~al.} 2011, \apj, 728, 122

\end{thebibliography}

\end{document}